\documentclass[a4paper,12pt]{article}

\usepackage[verbose=true]{microtype}
\usepackage{amsmath}
\usepackage{amssymb}
\usepackage{bbm}
\usepackage{mathrsfs}
\usepackage[perpage,symbol]{footmisc}
\usepackage{graphicx}
\usepackage[all,cmtip]{xy}
\usepackage{enumitem}
\usepackage[font=footnotesize]{caption}
\usepackage{authblk}
\usepackage[sort&compress,numbers]{natbib}
\usepackage{color}
\usepackage{quantum}
\usepackage{algpseudocode}
\usepackage[linktocpage,hypertexnames=false,pdfpagelabels]{hyperref}
\usepackage[amsmath,thmmarks,hyperref]{ntheorem}
\usepackage[capitalise,poorman]{cleveref}

\newtheorem{theorem}{Theorem}
\newtheorem{lemma}[theorem]{Lemma}
\newtheorem{corollary}[theorem]{Corollary}

\newtheorem{definition}[theorem]{Definition}

\theoremstyle{break}
\newtheorem{problem}[theorem]{Problem}

\theoremstyle{break}
\theorembodyfont{\normalfont}
\newtheorem{algorithm}[theorem]{Algorithm}

\theoremstyle{nonumberplain}
\theorembodyfont{\normalfont}
\theoremsymbol{$\Box$}
\newtheorem{proof}{Proof}

\crefname{equations}{Eqs.}{Eqs.}
\Crefname{equations}{Equations}{Equations}
\creflabelformat{equations}{\textup{(#2#1#3)}}
\crefrangelabelformat{equations}{\textup{(#3#1#4)}--\textup{(#5#2#6)}}
\crefname{condition}{Condition}{Conditions}

\newenvironment{instance}{\par\textbf{Instance:}}%
  {\par\noindent\ignorespacesafterend}
\newenvironment{question}{\par\noindent\textbf{Question:}}%
  {\par\noindent\ignorespacesafterend}
\newenvironment{promise}{\par\noindent\textbf{Promise:}}%
  {\par\noindent\ignorespacesafterend}

\renewcommand{\subparagraph}[1]{\noindent{\bf #1\\}}

\newcommand{\affilcr}{\protect\\}

\DeclareMathOperator{\dom}{dom}

\DeclareMathOperator{\dist}{dist}
\renewcommand{\vec}[1]{\mathbf{#1}}

\newcommand{\keyword}[1]{\textit{#1}}
\newcommand{\vv}{\vec{v}}
\newcommand{\xx}{\vec{x}}
\newcommand{\tL}{\tilde{L}}
\newcommand{\tA}{\tilde{A}}
\newcommand{\tdelta}{\tilde{\delta}}

\newcommand{\kket}[1]{\Vert #1\rangle}
\newcommand{\brakket}[2]{\langle #1\Vert #2\rangle}
\newcommand{\csproblem}[1]{\textsc{#1}}
\renewcommand{\order}[1]{{O}(#1)}

\newcounter{enumend}

\renewcommand{\paragraph}[1]{\vspace{0.5em}\textbf{\textit{#1}}\hspace{1em}}

\definecolor{grey}{rgb}{0.4,0.4,0.4}

\begin{document}

\title{The Complexity of Relating Quantum Channels to Master Equations}
\author[1,2]{Toby S. Cubitt\thanks{tcubitt@mat.ucm.es}}
\author[3,4]{Jens Eisert}
\author[5,6]{Michael M. Wolf}
\affil[1]{Department of Mathematics, University of Bristol\affilcr
  University Walk, Bristol BS8 1TW, UK}
\affil[2]{Departamento de An\'alisis Matem\'atico,
  Universidad Complutense de Madrid\affilcr
  Plaza de Ciencias 3, Ciudad Universitaria, 28040 Madrid, Spain}
\affil[3]{Dahlem Center for Complex Quantum Systems\affilcr
  Freie Universit{\"a}t Berlin, 14195 Berlin, Germany}
\affil[4]{Institute for Physics and Astronomy, Potsdam University,
  14476 Potsdam, Germany}
\affil[5]{Niels Bohr Institute, Blegdamsvej 17, 2100 Copenhagen, Denmark}
\affil[6]{Zentrum Mathematik, Technische Universit{\"at} M{\"u}nchen,
  85748 Garching, Germany}

\maketitle

\begin{abstract}
  Completely positive, trace preserving (CPT) maps and Lindblad master
  equations are both widely used to describe the dynamics of open quantum
  systems. The connection between these two descriptions is a classic
  topic in mathematical physics. One direction was solved by the now
  famous result due to Lindblad, Kossakowski Gorini and Sudarshan, who
  gave a complete characterisation of the master equations that generate
  completely positive semi-groups. However, the other direction has
  remained open: given a CPT map, is there a Lindblad master equation
  that generates it (and if so, can we find it's form)? This is sometimes
  known as the \keyword{Markovianity problem}. Physically, it is asking
  how one can deduce underlying physical processes from experimental
  observations.

  We give a complexity theoretic answer to this problem: it is NP-hard.
  We also give an explicit algorithm that reduces the problem to integer
  semi-definite programming, a well-known NP problem. Together, these
  results imply that resolving the question of which CPT maps can be
  generated by master equations is tantamount to solving P$=$NP: any
  efficiently computable criterion for Markovianity would imply P$=$NP;
  whereas a proof that P$=$NP would imply that our algorithm already
  gives an efficiently computable criterion. Thus, unless P does equal
  NP, there cannot exist any simple criterion for determining when a CPT
  map has a master equation description.

  However, we also show that if the system dimension is fixed (relevant
  for current quantum process tomography experiments), then our algorithm
  scales efficiently in the required precision, allowing an underlying
  Lindblad master equation to be determined efficiently from even a
  single snapshot in this case.

  Our work also leads to similar complexity-theoretic answers to a
  related long-standing open problem in probability theory.
\end{abstract}



\tableofcontents

\section{Introduction}
Noise abounds in quantum mechanical systems, so it's no surprise that the
mathematics of open quantum systems permeates many areas of quantum
theory. In quantum \emph{information} theory, noisy evolution is usually
modelled by completely positive, trace preserving (CPT) maps. CPT maps
are often referred to as \keyword{quantum channels}, as they play the
same role in quantum information theory as classical channels (stochastic
maps) play in classical information theory: they give a discrete,
black-box description of how input states are transformed into output
states.

Just as in classical information theory, questions ranging from
communication capacities to error-correction and fault-tolerant
computation benefit from abstracting away the underlying physics in this
way~\cite{Nielsen+Chuang}. CPT maps also arise naturally in experimental
measurement of quantum dynamics, when a complete ``snapshot'' of the
dynamics is reconstructed via \keyword{quantum process
  tomography}~\cite{Nielsen+Chuang}. The reconstructed snapshot is a CPT
map describing how initial states are transformed by the evolution into
states at the time of measurement.

Noisy evolution in other areas of quantum \emph{physics}, on the other
hand, is usually modelled by master equations. These directly describe
the underlying physical processes governing the evolution, in the form of
a differential equation for the time-evolution of the density matrix.
They are frequently used to model realistic experimental set-ups, where
external noise and dissipation must invariably be accounted for,
especially in quantum optics~\cite{Carmichael} and condensed-matter
physics~\cite{Weiss}.

In describing a noisy evolution by a master equation, there is an
implicit assumption that the effect of the external environment on the
system's evolution can be described in terms of the system's degrees of
freedom alone. Given this assumption, the master equation must
necessarily be \keyword{Markovian}. One justification for this is if the
underlying physical processes are forgetful---as they commonly are to a
good approximation. Conversely, if the Markovian assumption doesn't hold,
then there is no way to decribe the evolution physically without
enlarging the system being modelled to include (some of) the environment
degrees of freedom.

Mathematically, a Markovian master equation generates a one-parameter
(time $t$) semi-group (evolving for time $t$ and then time $s$ is
equivalent to evolving for time $t+s$) of CPT maps (the evolution must be
completely positive and trace preserving at all times if probabilities of
measurement outcomes are to be positive and sum to one).

\subsection{The Quantum Problem}
The connection between these two descriptions of open quantum
systems---the black-box, discrete-time description of CPT maps, and
the continuous-time, physical description of master equations---is a
classic topic in mathematical physics. Two questions naturally arise:
given a master equation, does it generate a a completely positive
evolution (and if so which CPT maps does it produce)? Conversely,
given one or more CPT maps, is there an underlying Markovian master
equation that generates them (and if so which one)? These questions
can equivalently be stated more mathematically: given a linear
operator, does it generate a completely positive semi-group?
Conversely, given one or more CPT maps, are they members of a
completely positive semi-group?

In seminal papers from the 1970's, Lindblad~\cite{Lindblad}, Gorini,
Koss\-ak\-ov\-ski and Sudarshan~\cite{Gorini} gave a complete answer
to the first question (for finite dimensional systems\footnote{For
  subtleties invovled in finding the most general form of a generator
  in infinite-dimensional quantum systems, see Ref.~\cite{Holevo}.}).
They derived the general form---now known as the \keyword{Lindblad
  form}---for the generators of one-parameter completely positive
semi-groups. Just as any discrete transformation of quantum states
must be completely positive and trace-preserving if probabilities are
to remain positive and normalised for any input state, a master
equation \emph{must} be of Lindblad form if it is to be physical,
since an evolution that is not of this form will necessarily lead to
negative probabilities.\footnote{There exists a large literature on
  ``non-Markovian master equations'', which are not of Lindblad form.
  These can provide a useful phenomenological description of quantum
  evolution. But since they necessarily predict negative probabilities
  for some physical measurement outcomes, they are only valid for a
  restricted set of ``allowed'' initial states. If the system is
  prepared in a state outside of this allowed set, the non-Markovian
  master equation becomes invalid.}

The converse question, however, has remained open. For the case of a
single CPT map, we will refer to the problem of deciding whether it is a
member of a completely positive semi-group as the \keyword{Markovianity
  problem}, since CPT maps that are generated by a Lindblad master
equation are said to be \keyword{Markovian}.\footnote{Note that this term
  is not used consistently throughout the literature. Here, we stick to
  the standard use of the term \keyword{Markovian} in the mathematical
  physics literature to mean the \keyword{time-homogeneous} Markovianity
  problem, in which the master equation is assumed to be
  time-independent. Sometimes, in particular in the context of
  condensed-matter physics, master equations are also referred to as
  being Markovian if they are of Lindblad form, but may be
  time-dependent. One could also adopt the established classical
  terminology and call the problem considered in this work the
  \keyword{quantum embedding problem}.} The main result of this work is a
complexity-theoretic answer to the Markovianity problem (which will be
made more rigorous later):
\begin{theorem}\label{thm:main}
  The Markovianity problem is NP-hard.
\end{theorem}
Our proof easily extends to more general problems, such as
deciding whether a family of CPT maps are members of the same completely
positive semi-group, or computing any ``measure'' of
Markovianity~\cite{markovianity,Cory,Cory2,Howard2,Weinstein,Lidar}.

``Hardness'' here is in the rigorous complexity-theoretic sense, which
will be explained more precisely below. (See also
Refs.~\cite{Papadimitriou,Computers+Intractability}.) It concerns the
scaling of computational effort as a function of the size of the problem,
i.e.\ as a function of the total amount of information required to
specify the CPT map. But a more refined analysis can break down the
overall problem size here into two components: the dimension of the
system, and the precision to which the CPT map is specified. We will
analyse the complexity of the Markovianity problem with respect to both
these parameters, and show that the NP-hardness is a consequence of
scaling of the dimension.\footnote{Note that the relevant parameter here
  is the system dimension, not the number of qubits (the base-2 logarithm
  of the dimension), as the amount of information required to specify the
  CPT map---the problem size---scales with the (square of) the dimension.
  The time required to perform process tomography scales only
  polynomially in the dimension, so is efficient in this context.} We
will also show---hinted at already in Ref.~\cite{markovianity}---that for
a \emph{fixed} dimension, the Markovianity problem can be decided
efficiently in the precision. Thus, though the problem in general is
(very likely) intractable, in practical contexts arising in current
quantum experiments, where the dimension is invariably small, the
question of whether a given (family of) CPT map(s) is consistent with
Markovian dynamics can be tested efficiently from even a single snapshot
in time. We will give an explicit algorithm in this case, along with a
careful analysis of its scaling:
\begin{theorem}\label{thm:algorithm}
  For any fixed physical dimension the Markovianity problem can be solved
  in a run-time that scales polynomially (both in the number of digits to
  which the entries of the CPT map are specified, and the precision to
  which the answer should be given).
\end{theorem}

\Cref{thm:main} proves that deciding Markovianity is at least as hard as
any problem in the complexity class NP. The algorithm of
\cref{thm:algorithm} reduces the problem to solving an integer
semi-definite program, a problem that is contained in the class NP.
Together, these results imply that:
\begin{corollary}
  Finding an efficiently computable criterion for Markovianity is
  equivalent to solving the (in)famous P$=$NP question; proving P$=$NP
  would imply the algorithm of \cref{thm:algorithm} is efficient, whereas
  finding any efficiently computable criterion for Markovianity would
  prove P$=$NP.
\end{corollary}

\subsection{The Classical Problem}
The analogous questions can equally well be posed for \emph{classical}
dynamics. In fact, the resulting mathematical problems are even older and
more extensively studied. The classical analogue of a CPT map is a
stochastic map, which, in the context of information theory, also
describes a classical communication channel. The classical analogue of a
master equation is a continuous-time Markov chain, and the Markov-chain
analogue of the Lindblad form can be found in any good text book on
Markov processes (see e.g.\ Ref.~\cite{Norris}).

However, the converse question: given a stochastic map, can it be
generated by a continuous-time Markov chain, has remained a thorny open
problem in probability theory for over 70 years! It is known as the
\keyword{embedding problem} for stochastic maps, and was first posed at
least as long ago as 1937 by Elfving~\cite{Elfving}. Though it has been
the subject of investigation over the many intervening
decades~\cite{Kingman,Kingman_Williams,Fuglede}, the general embedding
problem has remained open~\cite{Mukherjea} until now.

Although there is a sense in which the classical embedding problem can be
viewed as a special case of the quantum Markovianity problem,
mathematically the two are inequivalent: a result concerning one does not
necessarily imply anything about the other. However, it turns out that
very similar techniques can be used to tackle both problems, allowing us
to also show that:
\begin{theorem}
  The embedding problem is NP-hard.
\end{theorem}

This finally resolves the long-standing embedding problem, in the sense
that no efficiently computable (polynomial-time) criterion for
embeddability can exist unless P$=$NP; the existence of any such
efficiently computable criterion would imply P$=$NP\@. Rather than
duplicating everything for the classical case, we will focus in this
paper on the somewhat more complicated quantum problem, and then point
out how the results can be adapted to the older classical embedding
problem. A more detailed exposition of the classical result can be found
in Ref.~\cite{embedding}.

\subsection{Implications for Physics}
The Markovianity and embedding problems are not only of mathematical
interest. They are also crucial problems in physics. What is the best
possible measurement data that an experimentalist could conceivably
gather about a system's dynamics? They could, for example, repeatedly
prepare the system in any desired initial state, allow it to evolve for
some period of time, and then perform any desired measurement. In fact,
by choosing tomographically complete bases of initial states and
measurements, and carrying out this procedure only a finite number of
times, it is already possible to reconstruct a complete ``snapshot'' of
the system dynamics at any particular time to arbitrary accuracy. In the
quantum setting, this is \keyword{quantum process
  tomography}~\cite{Nielsen+Chuang}, but the general principle obviously
applies equally well in the classical setting. Remarkably, thanks to the
dramatic progress in experimental control and manipulation of quantum
systems over recent years, this is no longer a theoretical pipe-dream
even for quantum systems. Full quantum process tomography is now
routinely carried out in many different physical systems, from
NMR~\cite{TeleportNature,NMRreview,Cory,NMRScience} to trapped
ions~\cite{Blatt,BlattToffoli}, from photons~\cite{OBrien,Detectors}, to
solid-state devices~\cite{Howard}.

Each tomographic snapshot gives us a dynamical map, which tells us
\emph{everything} there is to know about the evolution at the time $t$
when the snapshot was taken. If, on the time scale of observation, the
discrete evolution is Markovian (i.e.\ doesn't depend on the history of
its past) then the snapshot determines how any initial state of the
system will evolve into a state at time $t$. This evolution is then
described mathematically by a stochastic map in the classical setting and
a CPT map in the quantum setting. In the quantum case, the indedepence
from the history, which is equivalent to having an uncorrelated joint
intial state of system and environment, can for instance be guaranteed if
the tomographic scheme can be carried out with pure input states. This is
certainly possible in principle, as we are assuming that the
experimentalist has full control over the initial state of the system,
and gives the best possible empirical description of the dynamics
accessible by an experiment. The quantum process tomography experiments
mentioned
above~\cite{TeleportNature,NMRreview,Cory,NMRScience,Blatt,BlattToffoli,OBrien,Detectors,Howard}
have carried this out to a good degree of approximation in a variety of
different physical systems.

Under this assumption, \emph{all} physical properties of the system at
time $t$ are then fully determined by the tomographic snapshot. In the
quantum case, the expectation value of \emph{any} physical observable $M$
is then given by Born's rule, whereas in the classical case it is given
by a straight-forward average. Any physical measurement can therefore be
viewed as an imperfect version of process tomography, since it gives
partial information about the snapshot, and with sufficient measurement
data the full snapshot can be reconstructed. Thus the most complete data
that can be gathered about a system's dynamics consists of a set of
snapshots, taken at different times during the evolution.

Given one or more snapshots, understanding the underlying physical
processes typically amounts to reconstructing the system's dynamical
equations and Liouvillian. If, over the time-scale of the experiment, the
dynamics is described to good approximation by Markovian dynamics, then
the dynamical equations take the form of a Lindblad master equation (in
the quantum case) or a continuous-time Markov process (in the classical
case). So to understand the physics underlying an experimental system, we
must understand whether they can be described by a Markovian dynamics,
and if so, what form the Markovian dynamical equations take. Clearly, if
we can \emph{find} a set of Markovian dynamical equations describing the
dynamics whenever these exist (and there is no a priori way of knowing
whether they exist or not), we can also determine \emph{whether} they
exist. So understanding the physics governing an experimental system
implicitly involves solving the Markovianity or embedding problem (or
their generalisations to a family of CPT or stochastic maps, in the case
of multiple snapshots).

Thus the results of this work have a surprising implication for physics:
no matter how much measurement data we might gather about the behaviour
of a physical system, deducing its underlying Markovian dynamical
equations---if the dynamics can be traced back to such a process---is
fundamentally an intractable problem (assuming P$\neq$NP). Indeed,
already deciding whether or not the Markov approximation is a reasonable
one given the experimental data is intractable. And this extends to
various closely related physical problems, such as finding the dynamical
equation that best approximates the data, or testing a dynamical model
against experimental data.

Given their importance to physics, it is not surprising that numerous
heuristic numerical techniques have been applied to tackle the
Markovianity and embedding
problems~\cite{Cory,Cory2,Howard2,Weinstein,Lidar}. But these methods
give no guarantee of finding the correct answer, or even any indication
as to whether the correct answer has been found. One implication of the
results of this work is that any such technique must necessarily fail in
the general case (although for fixed physical problem dimension, they can
of course prove valuable). The algorithm given in
\cref{sec:fixed_dimension}, which we prove is efficient for fixed
dimension, improves on previous methods in that it guarantees to give the
correct answer. It can also be extended to provide a similarly rigorous
\emph{measure} of the degree of Markovianity~\cite{markovianity}.


\subsection{Outline}
After introducing the necessary notation and recalling basic concepts in
\cref{sec:preliminaries}, \cref{sec:quantum} develops a careful and
rigorous formulation of the Markovianity problem that will allow us to
apply tools from complexity theory. \Cref{sec:NP-hardness} then gives a
complexity-theoretic answer to the Markovianity problem: it is NP-hard.
Technically, NP-hardness alone does not prove equivalence to P$=$NP; it
could be that the Markovianity problem is \emph{much} harder, so that
even P$=$NP would not imply an efficient algorithm for Markovianity.
\Cref{sec:fixed_dimension} completes the proof of equivalence by giving
an explicit algorithm that reduces the Markovianity problem to solving an
NP-complete problem. We give a careful analysis of the complexity of this
algorithm, thereby providing an explicit algorithmic solution to the
Markovianity problem which would be efficient if P$=$NP. Indeed, we show
that if the dimension is fixed, the algorithm scales polynomially in the
precision. In \cref{sec:classical} we briefly explain how these proofs
can be adapted to show that the classical embedding problem, too, is
NP-hard (a fuller version appears in Ref.~\cite{embedding}). Finally,
\cref{sec:conclusions} concludes with a discussion of consequences of
these results.

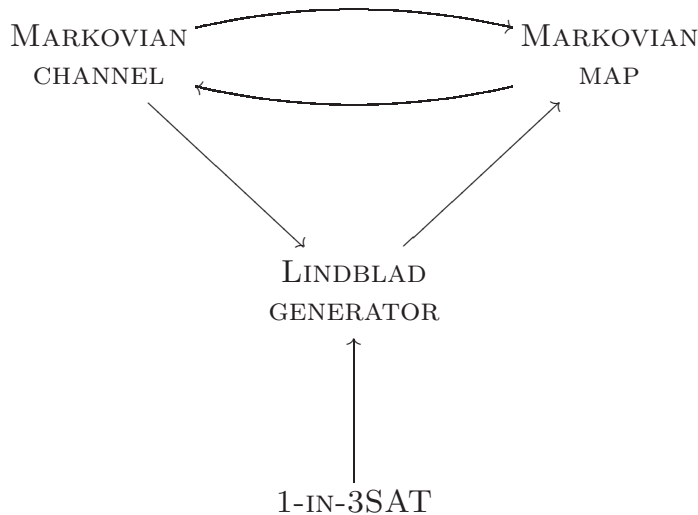
\begin{figure}[!htbp]
  \begin{displaymath}
    \xymatrix{
      {\begin{matrix}
          \text{\csproblem{Markovian}}\\
          \text{\csproblem{channel}}
       \end{matrix}}
      \ar[rdd] \ar@/^1.5pc/[rr]
      &&
      {\begin{matrix}
          \text{\csproblem{Markovian}}\\
          \text{\csproblem{map}}
       \end{matrix}}
      \ar@/^1.5pc/[ll]\\\\
      &{\begin{matrix}
          \text{\csproblem{Lindblad}}\\
          \text{\csproblem{generator}}
        \end{matrix}}
       \ar[ruu]
      \\\\
      &\text{\csproblem{1-in-3SAT}} \ar[uu]
    }
  \end{displaymath}
  \caption{Computational problems defined in the proof, along with the
    complexity-theoretic reductions between them. Since
    \csproblem{1-in-3SAT} is NP-complete, taken together this sequence of
    reductions proves NP-hardness of the Markovianity problem.}
  \label{fig:computational_problems}
\end{figure}

As the full NP-hardness proof described in
\cref{sec:quantum,sec:NP-hardness} is somewhat involved, we give here an
overview of the general structure of the argument, as an aid to
navigating the details of the proof. The proof proceeds by defining a
number of computational problems and proving a sequence of
complexity-theoretic relationships between them, starting from the
Markovianity problem itself, and ending with the NP-complete problem
\csproblem{1-in-3SAT}. The computational problems defined in the proof,
and the relationships we will establish between them, are illustrated in
\cref{fig:computational_problems}.

Just as the dynamics of a closed quantum system governed by a Hamiltonian
$H$ is described formally by a unitary semi-group $U_t=e^{Ht}$ obtained
by exponentiation, the dynamics of an open quantum system governed by a
Liouvillian $L$ of Lindblad form is described formally by a
completely-positive semi-group $E_t=e^{Lt}$ obtained by exponentiation of
the Liouvillian. However, unlike unitary dynamics, not every
completely-positive map can be generated by a Lindblad master equation.
The Markovianity problem is precisely the question of determining whether
a given CPT map $E$ \emph{is} generated by some Lindblad master equation
or not. In \cref{sec:quantum}, we formulate this question rigorously as
the computational problem \csproblem{Markovian channel}. This is the
first of our computational problems, and the one we are seeking to prove
is NP-hard.

It turns out to be helpful for the proof to define another variant of
this computational problem, called \csproblem{Markovian map}, in which
the map that we are given is not necessarily CPT. The first step in the
proof is to show that these two problems, \csproblem{Markovian channel}
and \csproblem{Markovian map}, are computationally equivalent; i.e.\
\csproblem{Markovian map} can be reduced to \csproblem{Markovian channel}
(the reduction in the opposite direction is trivial, since
\csproblem{Markovian channel} is just a special case of
\csproblem{Markovian map}). This is not difficult, and we do so at the
end of \cref{sec:computational_Markovianity}. This proves the first (and
simplest) of the complexity-theoretic relationships illustrated in
\cref{fig:computational_problems}.

For the finite-dimensional systems with which we are concerned, the
Liouvillian $L$ is given by a finte-dimensional matrix, and the
exponentiation $E_t=e^{Lt}$ is the standard matrix exponential. By
inverting this relationship e.g.\ for $t=1$, we obtain an expression for
the Liouvillian $L = \log E_1$ in terms of the matrix logarithm. In this
way, the Markovianity problem for CPT map $E$ becomes one of determining
whether $L = \log E$ is of Lindblad form. In
\cref{sec:computational_Lindblad}, we show that there is a simple and
computationally efficient algorithm for determining whether a given
matrix $L$ is of Lindblad form. The difficulty lies in the fact that the
logarithm $\log E$ is not uniquely defined. Just as there are infinitely
many logarithms $\log r + i\phi + 2\pi i n$ of a complex number
$z=re^{i\phi}$, parameterised by an integer $n\in\ZZ$, there are
infinitely many branches of the matrix logarithm, parameterised now by a
vector of integers. Thus to solve the Markovianity problem for a map $E$,
we must check whether \emph{any one} of the infinitely many possible
logarithms are of Lindblad form. In \cref{sec:computational_Lindblad}, we
formulate this rigorously as the computational \csproblem{Lindblad
  generator}.

It is worth pausing at this point to note that, already here, we see a
hint as to why the Markovianity problem might be NP-hard. In terms of the
Liouvillian, the problem is one of checking whether any element of a set
parameterised by integers (the possible logarithms) has a particular
property (the Lindblad form). There are of course many exceptions, but it
is often the case that integer problems such as this are NP-hard. For
example, linear programming problems can be solved efficiently, but
\emph{integer} linear programming is NP-complete. Indeed, it is trivial
to express NP-complete satisfiability problems such as \csproblem{3SAT}
as integer linear programs. Though the construction is significantly more
complicated, the same idea lies behind our NP-hardness proof for the
\csproblem{Lindblad generator} problem.

The remainder of \cref{sec:computational_Lindblad} is taken up with
proving that the \csproblem{Lindblad generator} problem is
computationally equivalent to \csproblem{Markovian channel}. In fact, we
first prove that \csproblem{Lindblad generator} can be reduced to
\csproblem{Markovian map}, implying that \csproblem{Markovian map} is
computationally least as difficult as \csproblem{Lindblad generator}.
Then we prove a reduction from \csproblem{Markovian channel} to
\csproblem{Lindblad generator}, implying that \csproblem{Lindblad
  generator} is computationally at least as hard as \csproblem{Markovian
  map}. Since we have already seen that \csproblem{Markovian channel} and
\csproblem{Markovian map} are computationally equivalent, this implies
equivalence of all three problems. This is illustrated in
\cref{fig:computational_problems}.

Having proven that the Markovianity problems are equivalent to the
\csproblem{Lindblad generator} problem, the final stage is to prove
NP-hardness of the latter. We do this in \cref{sec:NP-hardness} by
proving a reduction from a well-known NP-complete problem
\csproblem{1-in-3SAT} (a close cousin of the more famous \csproblem{3SAT}
problem), implying that the \csproblem{Lindblad generator} problem is at
least as hard as \csproblem{1-in-3SAT}. By the sequence of relationships
already proven between \csproblem{Lindblad generator} and
\csproblem{Markovian channel}, this implies NP-hardness of the
Markovianity problem. The complete sequence of relationships is
illustrated in \cref{fig:computational_problems}.

\section{Preliminaries}
\label{sec:preliminaries}
In what follows, we will restrict our attention to finite-dimensional
spaces and maps. It will be convenient to choose a concrete
representation for the CPT maps. Since a CPT map $\mathcal{E}$ is a
linear map on the $d^2$--dimensional vector space $\mathcal{M}_d$ of
operators on a $d$--dimensional Hilbert space $\mathcal{H}$, it can be
represented by a $d^2\times d^2$--dimensional matrix $E$ in the usual
way. More explicitly, if we reshape the density matrix $\rho$ as a vector
$\kket{\rho}$ with elements $\brakket{i,j}{\rho} = \rho_{i,j}$ in some
orthonormal basis, $E$ has matrix elements
\begin{equation}
  E_{(i,j),(k,l)} = \brakket{i,j}{\mathcal{E}(\ketbra{k}{l})}.
\end{equation}
The action of the channel $\mathcal{E}$ is then given by matrix
multiplication, $\kket{\mathcal{E}(\rho)} = E\kket{\rho}$, and the
composition $\mathcal{E}_1\circ\mathcal{E}_2$ of two channels
$\mathcal{E}_1$ and $\mathcal{E}_2$ is given in this linear operator
representation by the matrix product $E_1E_2$.

The matrix $E$ is also closely related to the more familiar
Choi-Jamio\l{}kowski state representation~\cite{Choi,Jamiolkowski}, given
by the state $\sigma = (\mathcal{E}\otimes\mathcal{I})(\omega)$ obtained
by applying the channel to one half of the (unnormalised) maximally
entangled state $\omega = \sum_{i,j}\ketbra{i,i}{j,j}$, defined in some
fixed orthonormal product basis of $\mathcal{M}_d\otimes\mathcal{M}_d$
($\mathcal{I}$ being the identity map). Define the involution $\Gamma$ by
its action on this basis,
\begin{equation}
	\ketbra{i,j}{k,l}^\Gamma = \ketbra{i,k}{j,l}.
\end{equation}
The Choi-Jamio\l{}kowski and linear operator representations of
$\mathcal{E}$ are then related by $E = \sigma^\Gamma$.

Completely positive semi-groups of CPT maps $\mathcal{E}_t$ arise
naturally as solutions of a Markovian quantum \keyword{master equation}
describing the dynamics of the density matrix $\rho$ (indeed, the
continuous semi-group structure is essentially the \emph{only} possible
one if we require the evolution to be describable at any time $t\geq
0$~\cite{divisibility,Denisov}):
\begin{equation}
  \frac{\dd\rho}{\dd t} = \mathcal{L}(\rho),
\end{equation}
where $\mathcal{L}$ is the system's Liouvillian. If the solutions
$\rho(t) = \mathcal{E}_t(\rho(0))$ are to be completely positive for all
$t\geq 0$, then the Liouvillian $\mathcal{L}$ must be of Lindblad
form~\cite{Lindblad,Gorini}:
\begin{equation}\label{eq:Lindblad_form}
  \frac{\dd\rho}{\dd t} = \mathcal{L}(\rho)
  = i[\rho,H] + \sum_{\alpha,\beta} G_{\alpha,\beta}\Bigl(
  F_\alpha\rho F_\beta^\dagger
  - \frac{1}{2}\{F_\beta^\dagger F_\alpha,\rho\}_+
  \Bigr).
\end{equation}
Here, $H$ is Hermitian, and can be interpreted as the Hamiltonian of the
system, $G\geq 0$ and $\{F_\alpha\}$ describe the decoherence processes,
and $\{A,B\}_+ = AB + BA$ denotes the anti-commutator. A
\keyword{Markovian} channel is one that is a member of such a semi-group,
i.e.\ one that is generated by some $\mathcal{L}$ of the above form.

It will again be convenient to represent the generator $\mathcal{L}$ by a
matrix, in the same way as for the channels. In the linear operator
representation, a Markovian channel $E = e^L$ is one with a generator $L$
such that $e^{Lt}$ is CPT for all $t\geq 0$. Note that we can without
loss of generality rescale time such that $E$ is generated by $L$ at time
$t=1$. The fact that the generator and channel are related by standard
matrix exponentiation in the linear operator representation makes this
representation particularly convenient for our purposes. It is not
difficult to translate \cref{eq:Lindblad_form} into conditions on $L$
(see \cref{sec:computational_Lindblad} or Ref.~\cite{markovianity}).

The classical case is analogous. A stochastic map on a finite
$d$--dimensional state space is represented by a $d\times d$--dimensional
stochastic matrix $P$, which acts on $d$--dimensional probability vectors
$\vec{p}$. An \emph{embeddable} stochastic matrix $P=e^Q$ is then one
with a generator $Q$ such that $e^{Qt}$ is stochastic for all $t\geq 0$,
i.e.\ $Q$ defines a continuous-time Markov chain. The conditions on $Q$
analogous to the Lindblad form of \cref{eq:Lindblad_form} (or, more
precisely, to \cref{lem:Lindblad}) are given by~\cite{Norris}:
\begin{enumerate}[labelindent=\parindent,leftmargin=*,
                  align=left,widest=iii]
  \item $Q_{i\neq j} \geq 0$ \label{cond:Q_positivity},
  \item $\sum_i Q_{i,j} = 0$ \label{cond:Q_normalisation}.
\end{enumerate}
For consistency with the quantum notation, we are adopting the convention
that probability distributions are \emph{column} vectors, and maps act on
them to the right. Thus the normalisation condition applies to the
column-sums rather than the row-sums. Note, however, that this runs
counter to the convention in the probability theory literature of
representing probability distributions by row-vectors.

We will also make use of some basic concepts from complexity theory. (See
e.g.\ Refs.~\cite{Papadimitriou,Computers+Intractability} for an
introduction to this field.) Complexity theory is concerned with how the
computational resources (typically time or space) required to solve a
problem scale with the problem size, where the size of a computational
problem is the amount of information required to specify the problem. The
most important complexity classes are defined for decision problems:
problems with ``yes'' or ``no'' answers. For example, the complexity
class P is defined as the class of all decision problems that can be
solved on a classical computer in a time that scales as a polynomial of
the problem size. We say that such problems can be solved in
\keyword{polynomial time}, or \keyword{efficiently}. The notorious
complexity class NP is defined as the class of all decision problems for
which, if the answer is ``yes'', there exists a proof that can be
\emph{verified} in polynomial time. Clearly, any problem in P is also in
NP. It is widely believed that P is a strict subset of NP; this is the
famous P versus NP problem, which remains open to this day. A classic
example of an NP problem that is not known to be in P is the
satisfiability problem: deciding whether there exists an assignment of
truth values to a set of boolean variables for which a given boolean
expression evaluates to ``true''. Finding such an assignment may be
difficult, but if such an assignment exists, then there clearly exists a
proof of this fact which can be evaluated efficiently: namely, the list
of truth assignments itself.

We say that a decision problem $A$ can be \keyword{reduced} to a decision
problem $B$ if there exists an algorithm that transforms any instance of
$A$ into an instance of $B$, such that the answer to this $B$ instance
gives the answer to the orignal $A$ instance. To give a meaningful
hierarchy of complexity classes, the computational resources allowed in
the reduction must be restricted in some way. For the complexity class
NP, the appropriate reductions are \keyword{polynomial-time reductions}.
\footnote{Strictly speaking, what we have described here is
  polynomial-time many-to-one reduction, or \keyword{Karp reduction}, the
  strongest form of reduction. This is the type of reduction used to
  define NP-hardness, and is the only form of reduction with which we
  will be concerned in this paper.} If $A$ has a polynomial-time
reduction to $B$, then $B$ is in a well-defined sense ``harder'' than
$A$, since an efficient algorithm for solving $B$ would also give an
efficient algorithm for $A$. Reduction defines a partial order on
computational problems, and we will write $A \leq B$ when $A$ has a
polynomial-time reduction to $B$. A problem $A$ is called
\keyword{NP-hard} if every problem in NP has a polynomial-time reduction
to $A$. An NP-hard problem that is also contained in NP is called
\keyword{NP-complete}. NP-complete problems are, in the above sense, the
hardest problems in NP.

\section{The Quantum Problem}
\label{sec:quantum}

\subsection{The Computational Markovianity Problem}
\label{sec:computational_Markovianity}
In order to apply tools from complexity theory to study the Markovianity
problem, we will need to define the problem in such a way that the
problem size---the amount of information needed to specify an instance of
the problem---is well-defined. Even in the finite-dimensional case, this
requires a little care. Since CPT maps form a continuous set, there may
exist Markovian and non-Markovian channels that are arbitrarily close (in
any distance measure). Thus, to guarantee an unambiguous answer in all
cases, the channel would need to be specified to infinite precision.

There are essentially two standard ways of dealing with this in
complexity theory. But, before we do so, it is instructive to first take
a step back and recall some of physical motivation for the problem. In
measuring a tomographic snapshot of a system's dynamics, there will
always be some experimental error, and it makes little sense to require
an answer that is more precise than this error. Mathematically, this
suggests that we should consider the Markovianity problem solved if we
can answer the question for some map that is a sufficiently close
approximation to the one we were given.

This is the intuitive idea behind the following \keyword{weak-membership}
formulation of the Markovianity problem (cf.\ Ref.~\cite{Gurvits}, which
uses a weak-member\-ship formulation of the separability problem):
\begin{problem}[MARKOVIAN CHANNEL]\label{prob:Markov_channel}
  \begin{instance}
    $(E,\varepsilon)$: CPT map $E$, precision $\varepsilon \geq 0$.
  \end{instance}
  \begin{question}
    Assert either that:
    \begin{itemize}
    \item for some map $E'$ with $\matnorm{E'-E}\leq\varepsilon$, there
      exists a map $L'$ such that $E'=e^{L'}$ and $e^{L't}$ is CPT for
      all $t\geq 0$;
    \item for some CPT map $E'$ with $\matnorm{E'-E}\leq\varepsilon$, no
      such $L'$ exists.
    \end{itemize}
  \end{question}
\end{problem}
Here, we do not specify the matrix norm $\Matnorm{.}$ in the problem
definition. However, given the equivalence of norms on finite-dimensional
spaces, with at most a polynomial prefactor in the dimension relating one
norm to the other, we can leave the choice of norm open for now. Again,
we can always without loss of generality scale time such that, if a
suitable $L'$ exists, $E'$ is generated by $L'$ at time $t=1$.

Note that, if $E$ is close to the boundary of the set of Markovian
channels, then it will be close to both Markovian and non-Markovian maps,
and both assertions will be valid simultaneously. The physical
interpretation in such a case would simply be that the snapshot was not
measured to sufficient precision to allow an unambiguous answer. (There
are other ways to formulate weak-membership problems, but they are
essentially equivalent~\cite{Ioannou}.) The other standard approach would
be to restrict $E$ to have rational entries, but this is less natural in
the present context.

Because there are cases in which both answers may be valid, the
weak-mem\-ber\-ship formulation of \csproblem{Markovian channel} is not
formally a decision problem. This by definition rules it out of the
decision class NP, where it by rights belongs. Whilst it is possible to
reformulate it as a decision problem, we will avoid getting bogged down
in these complexity theoretic technicalities here, and accept that
\csproblem{Markovian channel} is not in NP\@. (In fact, the appropriate
complexity class for weak membership problems is known as promise-NP,
which is like NP but with an additional promise that the problem instance
will never be in some set. The results of \cref{sec:fixed_dimension} show
that the Markovianity problem is indeed in promise-NP, which, together
with the NP-hardness result, implies that it is promise-NP-complete. See
Ref.~\cite{Ioannou} for a discussion of similar issues in the context of
the separability problem.)

\csproblem{Markovian channel} carries the implicit promise that $E$ is a
CPT map. It is natural to ask whether this affects the complexity of the
problem. After all, if a tomographic snapshot is measured experimentally,
it is very unlikely to be either precisely trace-preserving or completely
positive. This motivates the definition of the following variant of the
Markovianity problem, which accounts for non-CPT maps $E$:

\begin{problem}[MARKOVIAN MAP]\label{prob:Markov_map}
  \begin{instance}
    $(E,\varepsilon,\varepsilon')$: Map $E$, precision parameters
    $\varepsilon > \varepsilon' > 0$.
  \end{instance}
  \begin{question}
    Assert either that:
    \begin{itemize}
    \item for some map $E'$ with $\matnorm{E'-E}\leq\varepsilon$, there
      exists a map $L'$ such that $E'=e^{L'}$ and $e^{L't}$ is CPT for
      all $t\geq 0$;
    \item for some CPT map $E'$ with $\matnorm{E'-E}\leq\varepsilon$, no
      such $L'$ exists;
    \item no CPT map $E'$ exists for which $\matnorm{E'-E}\leq\varepsilon'$.
    \end{itemize}
  \end{question}
\end{problem}

It is not difficult to see that the two problems, \csproblem{Markovian
  channel} and \csproblem{Markovian map}, are in fact equivalent.
Clearly, \csproblem{Markovian channel} is a special case of
\csproblem{Markovian map}, in which the third assertion is always false
($E$ itself fulfils the requirements of one or other of the first two
assertions). Conversely, complete-positivity of a map $E$ is equivalent
to positivity of the Choi-Jamio\l{}kowski matrix $\rho = E^\Gamma$, and
$E$ is trace-preserving iff the partial trace of $\rho$ is the identity
matrix. So finding the closest CPT map $E'$ to $E$ is equivalent to
finding the closest positive-semi-definite, suitable matrix $\rho'$ to
$\rho$. Indeed, if we fix the norm in \csproblem{Markovian map} to be the
Frobenius norm\footnote{The Frobenius norm is convenient for two reasons:
  firstly, the square of the norm-distance $\matnormF{A-B}^2$ is strictly
  convex; secondly, it is invariant under permutation of matrix elements,
  in particular $\matnormF{A^\Gamma}=\matnormF{A}$.}
$\matnormF{A}:=(\sum_{i,j}A_{i,j}^2)^{1/2}$, then not only do we have
$\matnormF{E'-E} = \matnormF{\rho'-\rho}$, but also, if we minimise
$\matnormF{\rho'-\rho}^2$ subject to the above semi-definite constraints,
the objective function becomes a convex quadratic form. The problem can
therefore be transformed into a semi-definite program using standard
techniques~\cite{semidef}, allowing it to be solved efficiently to give
$E'$ and $\matnormF{E'-E}$. (More precisely, we can compute a bound on
$\matnormF{E'-E}$ that can be made exponentially tight with only
polynomial overhead.) Thus, either we will conclude that the third
assertion is valid, or we will succeed in transforming the problem into a
\csproblem{Markovian channel} instance. This proves the following
complexity-theoretic (Karp) equivalence\footnote{Throughout this paper,
  we will only consider Karp-reductions---i.e.\ polynomial-time
  reductions which transform one problem directly into a single instance
  of another---and Karp-equivalence. These are the strongest forms of
  reduction and equivalence, and are the ones used to define
  NP-hardness.}:
\begin{theorem}
  \label{thm:Markov_map-channel_equivalence}
  \csproblem{Markovian map} = \csproblem{Markovian channel}.
\end{theorem}

\subsection{The Computational Lindblad Generator Problem}
\label{sec:computational_Lindblad}
It is not immediately clear how one would go about solving a
\csproblem{Markovian channel} or \csproblem{Markovian map} instance. In
order to answer this, we will need to establish certain properties of the
generators $L$ of Markovian maps $E=e^{Lt}$. We will call such $L$
\keyword{Lindblad generators}. The following Lemma is taken directly from
Ref.~\cite{markovianity}, which in turn is a slight modification of the
argument given in Ref.~\cite{Lindblad}, and gives an efficient criterion
for deciding whether or not $L$ generates a one-parameter CPT semi-group,
i.e.\ whether it is of Lindblad form.
\begin{lemma}
  \label{lem:Lindblad}
  A map $L$ is a Lindblad generator iff all of the following hold:
  \begin{enumerate}[labelindent=\parindent,leftmargin=*,
                    align=left,widest=iii]
  \item $L^\Gamma$ is Hermitian. \label[condition]{cond:L_Hermiticity}
  \item $L$ fulfils the normalisation $\bra{\omega}L=0$, where the
    maximally entangled state vector $\ket{\omega}=\sum_i\ket{i,i}/
    \sqrt{d}$ is expressed in the same basis in which the involution
    $\Gamma$ is defined. \label[condition]{cond:L_normalisation}
  \item $L$ satisfies \label[condition]{cond:L_ccp}
    \begin{equation}\label{eq:ccp}
      (\id\matnorm-\omega)L^\Gamma(\id-\omega)\geq 0
    \end{equation}
    where $\omega=\proj{\omega}$.
  \end{enumerate}
\end{lemma}
Maps $L^\Gamma$ satisfying \cref{eq:ccp} are called
\keyword{conditionally completely positive} (ccp).

We can assume without loss of generality that the matrix $E$ in a
\csproblem{Markovian map} or \csproblem{Markovian channel} instance is
diagonalisable (with respect to similarity transforms), non-degenerate,
and full-rank. (Such matrices are dense in the set of all matrices, so we
can always replace $E$ with a neighbouring map that has these properties,
and decrease $\varepsilon$ (keeping $\varepsilon'$ fixed in the case of
\csproblem{Markovian map}) such that the outcome is unchanged.) The
Jordan decomposition of a diagonalisable channel has the form
\begin{equation}\label{eq:Jordan}
  E = \sum_r\lambda_r\Ketbra{r_r}{l_r}
      + \sum_c \lambda_c\Ketbra{r_c}{l_c}
      + \bar{\lambda}_c\,
        \mathbb{F}(\Ketbra{r_c}{l_c}).
\end{equation}
where $r$ labels the real eigenvalues, $c$ the complex ones, and
$\ketbra{r_k}{l_k}$ are orthonormal (but typically not self-adjoint)
spectral projectors formed from the left and right eigenvectors
$\bra{l_k}$ and $\ket{r_k}$ of $E$ associated with the same eigenvalue
$\lambda_k$. The fact that the eigenvalues come in conjugate pairs and
that the corresponding spectral projectors are related via the ``flip''
operation,
\begin{equation}
  \mathbb{F}\Bigl(\sum_{i,j}c_{i,j}\ket{i,j}\Bigr)
  = \sum_{i,j}\bar{c}_{i,j}\ket{i,j}
\end{equation}
extended to operators as
\begin{equation}
  \mathbb{F}\Bigl(
    \sum_{\mathclap{(i,j),(k,l)}}c_{(i,j),(k,l)}\ket{i,j}\bra{k,l}
  \Bigr)
  = \sum_{\mathclap{(i,j),(k,l)}} c_{(i,j),(k,l)}\ket{j,i}\bra{k,l},
\end{equation}
is a straightforward consequence of Hermiticity of CPT maps. It is easy
to show that all CPT maps are necessarily Hermitian.

Inverting the relationship $E=e^L$, we obtain a generator $L=\log E$ from
any channel $E$, where the matrix logarithm is defined via the logarithm
of the eigenvalues. Of course, the logarithm is not unique. It has a
countable infinity of branches, since the phase of each eigenvalue is
only determined modulo $2\pi$. $E$ is Markovian iff there exists
\emph{some} branch of the logarithm that has Lindblad form, i.e.\ that
satisfies \cref{lem:Lindblad}. So, to check if a channel is Markovian, we
must check whether \emph{any} branch of its logarithm has Lindblad form.

Some of the branches can be ruled out immediately, using the condition
that Lindblad generators must also be Hermitian maps
(\cref{cond:L_Hermiticity} from \cref{lem:Lindblad}), which imposes that
eigenvalues come in conjugate pairs. The remaining set of possible
Lindblad generators for $E$ can be parametrised by
\begin{equation}\label{eq:logs}
  L_m := \log E = L_0 + 2\pi i \sum_c m_c
         \big(\ketbra{l_c}{r_c}
           -\mathbb{F}(\ketbra{l_c}{r_c})
         \big)
      = L_0 + \sum_c m_c A_c,
\end{equation}
where $L_0$ is any fixed branch of the logarithm, e.g.\ the principle
branch (defined by taking the principle branch in the logarithm of each
eigenvalue), and each branch is characterised by a set of at most $d^2/2$
integers $m_c$ (one for each pair of complex eigenvalues). We introduce
the matrices $A_c$, defined by
\begin{equation}\label{eq:A_c}
    A_c := 2\pi i \bigl(\ketbra{l_c}{r_c} - \mathbbm{F}(\ketbra{l_c}{r_c})\bigr)
\end{equation}
for notational convenience.

The $A_c$ are fully determined by $L_0$, or, equivalently, by $E$. The
following lemma summarises those properties of $A_c$ and $L_0$ that are
easy to check, and follows immediately from the first two conditions of
\cref{lem:Lindblad,eq:logs,eq:A_c}:
\begin{lemma}
  \label{lem:L0+Ac_properties}
  If $L_m = L_0 + \sum_c m_cA_c$ parametrise the logarithms of a CPT
  map $E$ as in \cref{eq:logs}, then $L_0$ and $A_c$ necessarily satisfy
  the following properties:
  \begin{enumerate}[labelindent=\parindent,leftmargin=*,
                    align=left,widest=iii]
  \item $L_0$ and $A_c$ are simultaneously diagonalisable.
  \item $A_c$ are mutually orthogonal, rank-2 matrices with non-zero
    eigenvalues $\pm 2\pi i$.
  \item $L_0$ and $A_c$ satisfy the normalisation $\bra{\omega}L_0$ =
    $\bra{\omega}A_c = 0$. \label[condition]{cond:normalisation}
    \setcounter{enumend}{\the\value{enumi}}
  \item The two eigenvalues of $L_0$ corresponding to the non-zero
    eigenvalues of any $A_c$ form a conjugate pair.
  \item The right and left eigenvectors $\ket{r_{1,2}}$ and
    $\bra{l_{1,2}}$ associated with a conjugate pair of eigenvalues
    are related by $\ket{r_2} = \mathbbm{F}(\ket{r_1})$ and
    $\bra{l_2} = \mathbbm{F}(\bra{l_1})$.
  \end{enumerate}
  The last two properties of pairs of eigenvalues and eigenvectors can
  be stated more concisely as:
  \begin{enumerate}[labelindent=\parindent,leftmargin=*,
                    align=left,widest=iii,label=(\roman*')]
  \setcounter{enumi}{\the\value{enumend}}
  \item $L_0^\Gamma$ and $A_c^\Gamma$ are Hermitian matrices.
  \end{enumerate}
\end{lemma}
Together with the ccp condition of \cref{lem:Lindblad},
\begin{equation}
  (\id-\omega)L_0^\Gamma(\id-\omega)
  + \sum_c m_c(\id-\omega)A_c^\Gamma(\id-\omega)
  \geq 0,
\end{equation}
this gives a criterion for deciding whether $L_m = L_0 + \sum_c m_c A_c$
generates a CPT semi-group. Note that it is possible for $L_m$ to be ccp even
if $L_0$ is not.

The characterisation of Lindblad generators in \cref{lem:Lindblad}
motivates the definition of a new weak-membership problem:

\begin{problem}[LINDBLAD GENERATOR]\label{prob:Lindblad}
  \begin{instance}
    $(L_0,\delta)$: Map $L_0$, precision $\delta$.
  \end{instance}
  \begin{promise}
    There exists a map $L_0'$ with $\matnorm{L_0-L'_0}\leq f(\delta)$
    such that $e^{L'_0}$ is a quantum channel. ($f(\delta)$ is a strictly
    increasing function of $\delta$ which will be specified later.)
  \end{promise}
  \begin{question}
    Assert either that:
    \begin{itemize}
    \item for some map $L'_0$ with $\matnorm{L'_0 - L_0} \leq \delta$,
      there exists a set of integers $\{m_c\}$ such that $L'_m$ as defined in
      \cref{eq:logs} satisfies \cref{lem:Lindblad};
    \item for some map $L'_0$ where $e^{L'_0}$ is a quantum channel and
      $\matnorm{L'_0 - L_0} \leq \delta$, no such $L'_m$ exists.
    \end{itemize}
  \end{question}
\end{problem}
The bound $f(\delta)$ in the promise will be a somewhat complicated
monotonically increasing function of $\delta$ whose definition we defer
until later (see \cref{thm:Lindblad-Markov_reduction}), when it will make
more sense. But, essentially, the promise guarantees that $L_0$ is close
to the generator of \emph{some} CPT map. This definition of
\csproblem{Lindblad generator} might appear somewhat arbitrary. And
indeed it would be, were we interested in the problem of deciding
Lindblad form per se. (In that case, it would make more sense to replace
the promise by an extra assertion, analogous to the third assertion of
\csproblem{Markovian map}.) But we will only use \csproblem{Lindblad
  generator} as a stepping-stone to results concerning
\csproblem{Markovian channel} and \csproblem{Markovian map}, and the
above definition fulfils this purpose. In a slight abuse of terminology,
we will also refer to maps $L_0$ for which there exists an $L_m$
satisfying \cref{lem:Lindblad} as \keyword{Lindblad generators}, even if
$L_0$ itself is not of Lindblad form.

The preceding discussion suggests that \csproblem{Lindblad generator}
and\linebreak \csproblem{Markovian map} are equivalent. Clearly, the map
$E=e^{L_0}$ is Markovian iff there exists at least one $L_m$ satisfying
\cref{lem:Lindblad}. However, a little care is required in order to show
that the reductions in both directions can be performed efficiently. In
particular, we must show that appropriate precision parameters
$\varepsilon$ and $\delta$ can be computed efficiently, as well as
accounting for the fact that the exponential and logarithm can not be
computed to infinite precision. This will require strong continuity
properties of the matrix exponential and logarithm, and whilst these are
easily established in the case of the exponential, they are somewhat more
complicated to establish for the
logarithm.

A proof of Lipschitz continuity of the exponential can be found in
standard texts (see e.g. Ref.~\cite[Corollary~6.2.32]{Horn+Johnson_Topics}).
\begin{lemma}
  \label{lem:exp_continuity}
  For any matrices $A$ and $B$ and any matrix norm $\Matnorm{.}$
  \begin{equation}\label{eq:exp_continuity}
    \Matnorm{e^A-e^B}
    \leq \exp(\Matnorm{A})\exp(\Matnorm{A-B})\Matnorm{A-B}.
  \end{equation}
\end{lemma}
For the logarithm, we will need the following definition and theorems
from Refs.~\cite{Kato} and~\cite{Weilenmann}.
\begin{definition}
  \label{def:distances}
  For closed linear operators $A,B$ on a Banach space, define
  \begin{align}
    d(A,B) &= \max[\delta(A,B),\delta(B,A)], \\
    \delta_1(A,B) &= \sup_{0<\lambda\leq 1}\delta(\lambda A,\lambda B),\\
    d_1(A,B) &= \max[\delta_1(A,B),\delta_1(B,A)],
  \end{align}
  (taken directly from Refs.~\cite{Weilenmann,Kato}, following the
  notation of Ref.~\cite{Weilenmann}). $\delta(A,B)$ is Kato's $\delta$
  measure~\cite[IV.\S 2.4]{Kato}.\footnote{The distance-like measure $d$
    (which Kato calls $\hat{\delta}$) goes variously by the names
    ``gap'', ``aperture'' or ``opening''. Here,
    \begin{equation}
      \delta(A,B) = \sup_{x}\dist( (\vec{x} ,A \vec{x}) , G(B)),
    \end{equation}
    where $G(B)$ is the graph of $B$, and the supremum is taken over all
    $\vec{x}$ in the domain of $A$, normalised such that $\|\vec{x}^2\| +
    \|A\vec{x}\|^2=1$.  This distance-like measure generates the
    correspondingly named topology. This topology can equivalently be
    defined as the standard graph topology on the graphs of the
    operators.}
\end{definition}
Note that none of these measures obey the triangle inequality, so none
are proper distance measures (though they can readily be turned into
such; see Ref.~\cite[IV.\S 2.2,2.4]{Kato}). The following theorem shows
that, on bounded operators, the topology generated by $\delta$ is
equivalent to the norm topology of the Banach space (see~\cite[\S IV,
Theorems~2.13 and~2.14]{Kato}).
\begin{theorem}
  \label{thm:dist_norm_equivalence}
  If $A$ and $B$ are bounded operators on a Banach space with norm
  $\Matnorm{.}$, then
  \begin{equation}\label{eq:dist_norm_equivalence}
    d(A,B) \leq \Matnorm{A-B}
  \end{equation}
  and, if in addition $d(A,B) < ({1+\matnorm{A}^2})^{1/2}$,
  \begin{equation}\label{eq:norm_dist_equivalence}
    \Matnorm{A-B}
    \leq \frac{(1+\Matnorm{A}^2)\delta(A,B)}
      {1-(1+\Matnorm{A}^2)^{1/2}\delta(A,B)}.
  \end{equation}
\end{theorem}

Continuity of the logarithm can now be stated in terms of the
distance-like measures of \cref{def:distances}
(see~\cite[Theorem~3.1]{Weilenmann}).
\begin{theorem}
  \label{thm:log_uniform_continuity}
  If $A,B\in\mathscr{P}_1(M)$ are operators on a Banach space with
  norm $\matnorm{.}$, then for $M>0$
  \begin{equation}
    d_1(\log A,\log B) \leq 134(1+M^2)\delta_1(A,B),
  \end{equation}
  where $\mathscr{D} = \{ A \;|\; \dom A\text{ dense}\}$
  and
  \begin{equation}
    \mathscr{P}_1(M)
      = \{A\in\mathscr{D} \;|\; \lambda\in\rho(A) \text{ and }
          (1-\lambda)\Matnorm{R(\lambda, A)}\leq M
          \text{ for }\lambda\leq 0 \}
  \end{equation}
  are subsets of operators on the Banach space, $R(\lambda,A)$ is the
  resolvent of $A$, and $\rho(A)$ its resolvent set.
\end{theorem}

For the case of finite-dimensional Hilbert spaces that we are concerned
with here, $\mathscr{P}_1(M)$ becomes the set of complex matrices whose
eigenvalues do \emph{not} lie on or close to the negative real axis. This
amounts to taking the branch-cut of the logarithm to be along that axis.
(Since this rules out zero eigenvalues, these matrices are also
necessarily non-singular.)

Because we defined our computational problems in terms of norm-distance,
rather than the distance-like measures of \cref{def:distances}, we need
to transform \cref{thm:log_uniform_continuity} into a statement about
norm-distance.
\begin{corollary}
  \label{cor:log_continuity}
  If $A,B$ are bounded operators on a Banach space with norm
  $\matnorm{.}$, and if $kA,kB\in\mathscr{P}_1(M)$ with
  \begin{equation}
    k = \min\left[1,
          \left(134^2(1+M)^2\matnorm{A-B}^2-\matnorm{A}^2\right)^{1/2}
        \right],
  \end{equation}
  then
  \begin{multline}
    \label{eq:log_continuity}
    \Matnorm{\log A - \log B}\\
    \leq 134k(1+M^2)\left(
        1+k\matnorm{A}+k\matnorm{A-B}({1+k^2\matnorm{A}^2})^{1/2}
      \right)\Matnorm{A-B}.
  \end{multline}
\end{corollary}
\begin{proof}
  Assume first that $d(\log A,\log B)<({1+\matnorm{\log A}^2})^{1/2}$, so
  that the condition of \cref{thm:dist_norm_equivalence} holds and
  \cref{eq:norm_dist_equivalence} is valid. From \cref{def:distances},
  and rearranging \cref{eq:norm_dist_equivalence}, we have
  \begin{equation}\label{eq:d1_gt_normlog}
    \begin{aligned}
      d_1(\log A,\log B)
      &\geq \delta_1(\log B,\log A)
      = \sup_{0<\lambda\leq 1}\delta(\lambda\log B,\lambda\log A)\\
      &\geq \delta(\log B,\log A)
      \geq \frac{\Matnorm{\log A - \log B}}
                {1+\Matnorm{A}+\Matnorm{A-B}(1+\Matnorm{A}^2)^{1/2}}
    \end{aligned}
  \end{equation}
  and
  \begin{equation}\label{eq:delta1_lt_norm}
    \begin{aligned}
      \delta_1(A,B)
      &= \sup_{0<\lambda\leq 1}\delta(\lambda A,\lambda B)
      \leq \sup_{0<\lambda\leq 1}d(\lambda A,\lambda B)\\
      &\leq \sup_{0<\lambda\leq 1}\Matnorm{\lambda A - \lambda B}
      =\Matnorm{A-B}.
    \end{aligned}
  \end{equation}
  Using these inequalities in \cref{thm:log_uniform_continuity} gives
  \cref{eq:log_continuity} of the Corollary with $k=1$, under the
  assumption that $d(\log A,\log B)$ obeys the condition of
  \cref{thm:log_uniform_continuity}.

  Otherwise, we can rescale $A$ and $B$ until they do obey the
  condition. Let
  \begin{equation}
    0 < k < \left(134^2(1+M^2)^2\matnorm{A-B}^2
                  - \matnorm{A}^2\right)^{-1/2}.
  \end{equation}
  Then, using \cref{eq:delta1_lt_norm} and
  \cref{thm:log_uniform_continuity},
  \begin{align*}
    d\bigl(\log(kA),\log(kB)\bigr)
    &\leq d_1\bigl(\log(kA),\log(kB)\bigr)
    \leq 134(1+M^2)\delta_1(kA,kB)\\
    &\leq 134\abs{k}(1+M^2)\matnorm{A-B}
    < ({1+\abs{k}^2\matnorm{A}^2})^{1/2}\\
    &= ({1+\matnorm{kA}^2})^{1/2},
  \end{align*}
  so $d\bigl(\log(kA),\log(kB)\bigr)$ does satisfy the condition of
  \cref{thm:log_uniform_continuity}, and by the preceding argument
  \cref{eq:log_continuity} applies to
  $\matnorm{\log(kA)-\log(kB)}$. But
  \begin{equation}
    \begin{aligned}
      \matnorm{\log(kA)-\log(kB)}
      &= \matnorm{\log A + \log(k\id) - \log B - \log(k\id)}\\
      &= \matnorm{\log A - \log B},
    \end{aligned}
  \end{equation}
  which completes the proof.
\end{proof}
Note that if $A$ or $B$ happens to have an eigenvalue on the negative
real axis, we can always rotate the branch-cut, or equivalently the
eigenvalues. Multiplying by a scalar root of unity $z$ rotates the
eigenvalues away from the real axis, without changing the bound in
\cref{cor:log_continuity}: $\matnorm{\log(zA) - \log(zB)}
= \matnorm{\log A - \log B}$, but $\matnorm{zA - zB}
= \matnorm{A-B}$.

We are now in a position to prove the main results of this section.
\begin{theorem}
  \label{thm:Lindblad-Markov_reduction}
  \csproblem{Markovian map} $\geq$ \csproblem{Lindblad generator}.
\end{theorem}
\begin{proof}
  Assume first that we are given an instance $(L_0,\delta)$ of
  \csproblem{Lindblad generator} that is unambiguous, i.e.\ either all
  neighbouring generators of channels are Lindblad generators, or none
  are. In that case we know that one or other of the assertions is valid,
  but not both. Now, using \cref{cor:log_continuity}, we can calculate
  (efficiently) an $\varepsilon$ such that for $\log E=L_0$, $\log E'=L'_0$,
  and $\matnorm{E-E'}\leq\varepsilon$, we have $\matnorm{\log E-\log
    E'}\leq\delta$. (Indeed, it is not difficult to solve
  \cref{eq:log_continuity} for $\varepsilon$ and obtain an explicit
  expression.) Then the pre-image of an $\varepsilon$-ball around
  $E=e^{L_0}$ is contained within the $\delta$-ball around $L_0$ (as
  illustrated in \cref{fig:epsilon-delta}). Since a map $E'=e^{L'_0}$ is
  Markovian iff $L'_0$ is a Lindblad generator, and we are assuming the
  \csproblem{Lindblad generator} instance is unambiguous, any channels
  within this $\varepsilon$-ball must either all be Markovian or all be
  non-Markovian.
  \begin{figure}[htb]
    \begin{center}
      \input{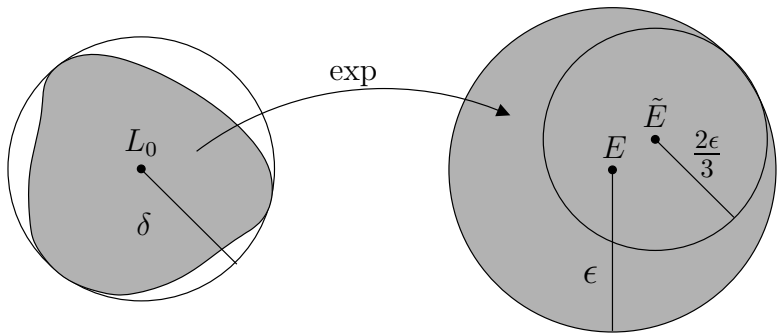}
    \end{center}
    \caption{The pre-image of an $\varepsilon$-ball around $E=e^{L_0}$ is
      contained within a $\delta$-ball around $L_0$. If $\tilde{E}$ is
      within $\varepsilon/3$ of $E$, then everything within a
      $2\varepsilon/3$-ball around $\tilde{E}$ is within the
      $\varepsilon$-ball around $E$.}
    \label{fig:epsilon-delta}
  \end{figure}

  To deal with the fact that $E=e^{L_0}$ can not be calculated to
  infinite precision, let $\tilde{E}$ be the exponential of $L_0$
  calculated to within precision $\varepsilon/3$ (which can be done
  efficiently~\cite{matexp}); i.e.\
  $\matnorm{\tilde{E}-E}\leq\varepsilon/3$. If $E'$ is within a
  $2\varepsilon/3$-ball around $\tilde{E}$, we have $\matnorm{E'-E} \leq
  \varepsilon$. Therefore, assuming for the moment that there exists some
  channel within this ball (i.e.\ assuming its third assertion is
  \emph{not} valid), the \csproblem{Markovian map} instance
  $(\tilde{E},2\varepsilon/3,\varepsilon')$ with any
  $\varepsilon'\leq2\varepsilon/3$ will return its first (second)
  assertion iff the first (second) assertion of the original
  \csproblem{Lindblad generator} instance was valid (always under the
  assumption that the original \csproblem{Lindblad generator} instance
  was unambiguous). This is illustrated in \cref{fig:epsilon-delta}.

  We must now justify the assumption that the third assertion of the
  \csproblem{Markovian map} instance $(\tilde{E},2\varepsilon/3,\varepsilon')$
  is always false. Recall that the \csproblem{Lindblad generator} promise
  guarantees existence of a generator $L'_0$ of a quantum channel within
  an $f(\delta)$-ball around $L_0$. For the assumption to be justified,
  this must imply existence of at least one quantum channel within an
  $\varepsilon'$-ball around $\tilde{E}$. We now take $f(\delta)$ to be
  defined implicitly using \cref{lem:exp_continuity}, such that for
  $\matnorm{L_0-L'_0}\leq f(\delta)$ we have
  $\matnorm{e^{L_0}-e^{L'_0}}\leq\varepsilon/3$. (Once again, substituting
  the explicit expression for $\varepsilon$ into \cref{eq:log_continuity}
  and solving for $f(\delta)$ would give an explicit definition for the
  latter, if so desired.) Then $\matnorm{\tilde{E}-E'} \leq 2\varepsilon/3$,
  so that $E'$ fulfils the requirements with
  $\varepsilon'=2\varepsilon/3$. \Cref{fig:epsilon-delta2} illustrates this.
  \begin{figure}[htb]
    \begin{center}
      \input{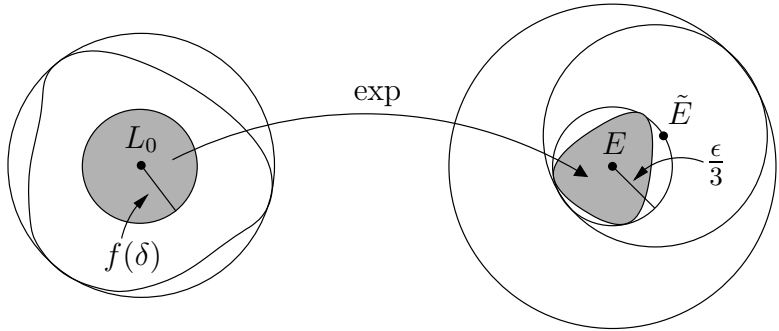}
    \end{center}
    \caption{Everything within an $f(\delta)$-ball around $L_0$ is
      mapped into an $\varepsilon/3$-ball around $E$, which itself is
      contained within a $2\varepsilon/3$-ball around $\tilde{E}$. (See
      also \cref{fig:epsilon-delta}.)}
    \label{fig:epsilon-delta2}
  \end{figure}

  Finally, it remains to consider the case of \csproblem{Lindblad
    generator} instances that \emph{are} ambiguous; i.e.\ there exist
  generators of both Markovian and non-Markovian channels within a
  $\delta$-ball around $L_0$. In that case, the \csproblem{Markovian map}
  instance $(\tilde{E},2\varepsilon/3,\varepsilon'=2\varepsilon/3)$ could
  return either assertion. But the original \csproblem{Lindblad
    generator} instance is also allowed to return either assertion in
  this case, which completes the proof of the reduction.
\end{proof}

\begin{theorem}
  \label{thm:Markov-Lindblad_reduction}
  \csproblem{Lindblad generator} $\geq$ \csproblem{Markovian channel}.
\end{theorem}
\begin{proof}
  The reduction from \csproblem{Markovian channel} to \csproblem{Lindblad
    generator} is very similar to the proof of
  \cref{thm:Lindblad-Markov_reduction}, reversing the roles of
  \cref{lem:exp_continuity} and \cref{cor:log_continuity}. The
  \csproblem{Lindblad generator} promise is automatically fulfilled, since
  $L_0=\log E$ is itself necessarily a generator of a quantum channel
  (namely, $E$).
\end{proof}
Together,
\cref{thm:Markov_map-channel_equivalence,thm:Lindblad-Markov_reduction,thm:Markov-Lindblad_reduction}
imply the following corollary:

\begin{corollary}
  \label{cor:Lindblad-Markov_equivalence}
  \csproblem{Lindblad generator} = \csproblem{Markovian map} =
  \csproblem{Markovian channel}.
\end{corollary}

\section{NP-hardness}
\label{sec:NP-hardness}
We are now in a position to consider the computational complexity of the
problems defined in the previous sections. Although the ccp condition of
\cref{eq:ccp} is an integer semi-definite program, and it is well known
that even linear integer programming is NP-complete, this by no means
proves that \csproblem{Lindblad generator} is NP-hard. Linear programming
is the special case of semi-definite programming in which the coefficient
matrices are diagonal. But the matrices $L_0$ and $A_c$ defining a
\csproblem{Lindblad generator} instance must satisfy a number of highly
non-trivial constraints, as listed in \cref{lem:L0+Ac_properties}, which
certainly cannot be satisfied by diagonal matrices. Instead, our approach
will be to restrict to a special case of \csproblem{Lindblad generator},
for which the relation between $L_0$ and $L_0^\Gamma$ is somewhat easier
to analyse, then show that this special case can be used to encode
\csproblem{1-in-3SAT}, a standard NP-complete satisfiability
problem~\cite{Computers+Intractability}, simpler even than its
better-known cousin 3SAT in that it does not require any boolean
negation:\footnote{Note that the use of the term 1-in-3SAT is not
  entirely consistent in the literature. Here we mean the variant that
  does not involve any negation, as originally formulated in
  Ref.~\cite{Schaefer}.}

\subsection{Encoding \csproblem{1-in-3SAT}}
\begin{problem}[1-in-3SAT]
  \begin{instance}
    $(n_v,n_C)$: $n_v$ boolean variables; $n_C$ clauses each with exactly
    3~variables.
  \end{instance}
  \begin{question}
    Is there a truth assignment of the variables such that each clause
    contains exactly one true variable?
  \end{question}
\end{problem}

\csproblem{1-in-3SAT} can be transformed into a set of simultaneous
linear integer inequalities in the standard way. Identify each boolean
variable with an integer variable $m_c$, and identify the values 1 and 0
with ``true'' and ``false''. For each $m_c$, write the inequalities
\begin{equation}\label[equations]{eq:boolean_constraints}
  m_c \geq -\frac{1}{2},\quad -m_c\geq -\frac{7}{6},
\end{equation}
and for each \csproblem{1-in-3SAT} clause involving variables $i$, $j$
and $k$, write the following inequalities:
\begin{equation}\label[equations]{eq:clause_constraints}
  m_i + m_j + m_k \geq \frac{1}{2},\quad
  -m_i - m_j - m_k \geq -\frac{3}{2}.
\end{equation}
The non-integer constants are chosen for later convenience. These
inequalities are satisfied for integer $m_c$ if precisely one $m_i$ from
each clause is equal to one and the others are all zero.

We now restrict the matrices $L_0$ and $A_c$ that define a \csproblem{Lindblad
  Generator} instance (cf.\ \cref{eq:logs}) to have the following special
forms:
\begin{align}
  L_0 &= \sum_{i,j} Q_{i,j}\ketbra{i,i}{j,j}
         + \sum_{i\neq j}P_{i,j}\ketbra{i,j}{i,j},
         \label{eq:special_case_L0}\\
  A_c &= 2\pi\sum_{i\neq j} B^c_{i,j}\ketbra{i,i}{j,j},
         \label{eq:special_case_Ac}
\end{align}
with
\begin{align}
  Q &= \sum_r\xx_r^{\vphantom{T}}\xx_r^T
         \otimes\begin{pmatrix}1&1\\1&1\end{pmatrix}
         \otimes\begin{pmatrix}
           k+\lambda_r&\lambda_r\\\lambda_r&k+\lambda_r
         \end{pmatrix}
         \notag\\
       &\qquad
       +\sum_{c}\vv_c^{\vphantom{T}}\vv_c^T\otimes
         \begin{pmatrix}\phantom{-}1&-1\\-1&\phantom{-}1\end{pmatrix}
         \otimes\begin{pmatrix}
           k&-\frac{1}{3}\\\frac{1}{3}&\phantom{-}k
         \end{pmatrix}
         \label{eq:special_case_Q}\\
       &\qquad
       +\sum_{c'}\vv_{c'}^{\vphantom{T}}\vv_{c'}^T\otimes
         \begin{pmatrix}\phantom{-}1&-1\\-1&\phantom{-}1\end{pmatrix}
         \otimes\begin{pmatrix}k&0\\0&k\end{pmatrix}\notag,\\
  B^c &= \vv_c^{\vphantom{T}}\vv_c^T\otimes
         \begin{pmatrix}\phantom{-}1&-1\\-1&\phantom{-}1\end{pmatrix}
         \otimes
         \begin{pmatrix}\phantom{-}0&1\\-1&0\end{pmatrix}.
         \label{eq:special_case_Bc}
\end{align}
$\{\xx_r\}$ and $\{\vv_c,\vv_{c'}\}$ are two complete sets of
mutually-orthogonal, real vectors, whilst $k$ and $\lambda_r$ are real.
Note that $Q$ and $B^c$ are normal matrices, as are $L_0$ and $A_c$.
Since $[L_0, A_c^\dagger]=0$, the $\{L_m = L_0 + \sum_c m_cA_c\}$ are
also normal. The factor of $2\pi$ in \cref{eq:special_case_Ac} is for
later convenience. \Cref{fig:L0,fig:Ac} give a graphical representation
of the structure of $L_0$ and $A_c$.

\begin{figure}[!htbp]
  \centering
  \color{black}
  \begin{equation*}
    \begin{pmatrix}
      \boxed{\mspace{10mu}\mathclap{Q_{1,1}}\mspace{10mu}%
        \rule[-1mm]{0mm}{4mm}} & &
      \boxed{\mspace{10mu}\mathclap{Q_{1,2}}\mspace{10mu}%
        \rule[-1mm]{0mm}{4mm}} & &
      \boxed{\mspace{10mu}\mathclap{Q_{1,3}}\mspace{10mu}%
        \rule[-1mm]{0mm}{4mm}} & \cdots \\
      &\rotatebox[origin=cb]{-45}{$\mathclap{%
          \underset{\rotatebox{45}{$P$}}{\boxed{\mspace{60mu}}}}$}
      \rule[-8mm]{0mm}{13mm}\mspace{15mu} & & & \\
      \boxed{\mspace{10mu}\mathclap{Q_{2,1}}\mspace{10mu}%
        \rule[-1mm]{0mm}{4mm}} & &
      \boxed{\mspace{10mu}\mathclap{Q_{2,2}}\mspace{10mu}%
        \rule[-1mm]{0mm}{4mm}} & &
      \boxed{\mspace{10mu}\mathclap{Q_{2,3}}\mspace{10mu}%
        \rule[-1mm]{0mm}{4mm}} & \cdots \\
      &&&\rotatebox[origin=cb]{-45}{$\mathclap{%
          \underset{\rotatebox{45}{$P$}}{\boxed{\mspace{60mu}}}}$}
      \rule[-8mm]{0mm}{13mm}\mspace{15mu} & \\
      \boxed{\mspace{10mu}\mathclap{Q_{3,1}}\mspace{10mu}%
        \rule[-1mm]{0mm}{4mm}} & &
      \boxed{\mspace{10mu}\mathclap{Q_{3,2}}\mspace{10mu}%
        \rule[-1mm]{0mm}{4mm}} & &
      \boxed{\mspace{10mu}\mathclap{Q_{3,3}}\mspace{10mu}%
        \rule[-1mm]{0mm}{4mm}} & \cdots \\
      \vdots & & \vdots & & \vdots & \ddots\\
    \end{pmatrix}
    \cong
    \begin{pmatrix}
      \boxed{\mspace{50mu}Q\rule[-10mm]{0mm}{22mm}\mspace{50mu}}&\\
      &\mspace{30mu}
       \rotatebox[origin=cb]{-45}{$\mathclap{%
          \underset{\rotatebox{45}{$P$}}{\boxed{\mspace{130mu}}}}$}
          \rule[-15mm]{0mm}{26mm}\mspace{50mu}\\
    \end{pmatrix}
  \end{equation*}
  \caption{The structure of $L_0$ from \cref{eq:special_case_L0} is most
    apparent if we reorder the rows and columns so that all the
    $(i,i),(j,j)$ elements are in the top, left corner. We can then think
    of $L_0\cong Q\oplus\diag P$ as being composed of a matrix $Q$ and a
    vector $P$.}
  \label{fig:L0}

  \begin{equation*}
    \begin{pmatrix}
      \boxed{\mspace{10mu}\mathclap{B^c_{1,1}}\mspace{10mu}%
        \rule[-1mm]{0mm}{4mm}} & &
      \boxed{\mspace{10mu}\mathclap{B^c_{1,2}}\mspace{10mu}%
        \rule[-1mm]{0mm}{4mm}} & &
      \boxed{\mspace{10mu}\mathclap{B^c_{1,3}}\mspace{10mu}%
        \rule[-1mm]{0mm}{4mm}} & \cdots \\
      &\phantom{\rotatebox[origin=cb]{-45}{$\mathclap{%
          \underset{\rotatebox{45}{$P$}}{\boxed{\mspace{60mu}}}}$}}
      \rule[-8mm]{0mm}{13mm}\mspace{15mu} & & & \\
      \boxed{\mspace{10mu}\mathclap{B^c_{2,1}}\mspace{10mu}%
        \rule[-1mm]{0mm}{4mm}} & &
      \boxed{\mspace{10mu}\mathclap{B^c_{2,2}}\mspace{10mu}%
        \rule[-1mm]{0mm}{4mm}} & &
      \boxed{\mspace{10mu}\mathclap{B^c_{2,3}}\mspace{10mu}%
        \rule[-1mm]{0mm}{4mm}} & \cdots \\
      &&&\phantom{\rotatebox[origin=cb]{-45}{$\mathclap{%
          \underset{\rotatebox{45}{$P$}}{\boxed{\mspace{60mu}}}}$}}
      \rule[-8mm]{0mm}{13mm}\mspace{15mu} & \\
      \boxed{\mspace{10mu}\mathclap{B^c_{3,1}}\mspace{10mu}%
        \rule[-1mm]{0mm}{4mm}} & &
      \boxed{\mspace{10mu}\mathclap{B^c_{3,2}}\mspace{10mu}%
        \rule[-1mm]{0mm}{4mm}} & &
      \boxed{\mspace{10mu}\mathclap{B^c_{3,3}}\mspace{10mu}%
        \rule[-1mm]{0mm}{4mm}} & \cdots \\
      \vdots & & \vdots & & \vdots & \ddots\\
    \end{pmatrix}
    \cong
    \begin{pmatrix}
      \boxed{\mspace{50mu}B^c\rule[-10mm]{0mm}{22mm}\mspace{50mu}}&&&&\\
      & 0 &   &      &   \\
      &   & 0 &      &   \\
      &   &   &\ddots&   \\
      &   &   &      & 0 \\
    \end{pmatrix}
  \end{equation*}
  \caption{Reordered in the same way, $A_c$ from
    \cref{eq:special_case_Ac} is composed of just a matrix part: $A_c
    \cong B^c\oplus 0$.}
  \label{fig:Ac}
\end{figure}

It is a simple matter to verify that the properties required by
\cref{lem:L0+Ac_properties} are indeed satisfied by the forms given in
\cref{eq:special_case_L0,eq:special_case_Ac,eq:special_case_Q,eq:special_case_Bc},
as long as
\begin{equation}
  \vec{w}^TQ = 0,
\end{equation}
and $P=P^\dagger $ is Hermitian, where $\vec{w}=(1,1,\dots,1)^T/\sqrt{d}$
for $d\times d$-matrix $Q$. Furthermore, the ccp condition of
\cref{lem:Lindblad} reduces to the pair of conditions
\begin{subequations}
\begin{gather}
  2\pi \sum_c  B^c_{i,j}\,m_c + Q_{i,j} \geq 0 ,\qquad i\neq j,
  \label{eq:special_case_ccp1}\\
    (\id-\vec{w}\vec{w}^T)K(\id-\vec{w}\vec{w}^T) \geq 0,
  \label{eq:special_case_ccp2}
\end{gather}
\end{subequations}
where $K$ denotes the $d\times d$-dimensional matrix with diagonal
elements $K_{i,i}=Q_{i,i}$ and off-diagonal elements $K_{i\neq
  j}=P_{i,j}$.

We encode the \csproblem{1-in-3SAT} inequalities of
\cref{eq:boolean_constraints,eq:clause_constraints} by writing them
directly into the $\{\vv_c\}$. We associate a single $\vv_c$ to each
boolean variable of the problem.  For each clause $l$, write a ``1'' in
the $l$'th element of the three $\vv_c$'s corresponding to the variables
appearing in that clause, and write a ``0'' in the same element of all
the other $\vv_c$. Since there are $n_C$ clauses in total, at the end of
this process the vectors each have $n_C$ elements. Now for each $\vv_c$,
write a ``1'' in its $n_C+c$'th element, writing a ``0'' in the
corresponding element of all the other vectors. So far, we have defined
the first $n_C+n_v$ elements of the vectors. Finally, extend the vectors
so that they are mutually orthogonal and all have the same Euclidean norm
$\vv_c^T\vv_c^{\vphantom{T}}$. This can always be done, and will require
at most a further $n_v$ elements, producing vectors with at most
$n_C+2n_v$ elements. This procedure encodes the coefficients for the
\csproblem{1-in-3SAT} inequalities into some of the on-diagonal $4\times
4$ blocks of the $B^c$. Specifically, if we imagine colouring $B^c$ in a
chess-board pattern (starting with a ``white square'' in the top-leftmost
element), then the coefficients for one inequality are duplicated in all
the ``black squares'' of one $4\times 4$ block (see
\cref{fig:Bc_clause_block}).

\begin{figure}[!htbp]
  \centering
  \color{black}
  \begin{equation*}
    B^i,B^j,B^k =
    \begin{pmatrix}
      \\
      &\ddots&&&&&&&\vline\\
      &\rule[-2mm]{0mm}{9mm}
      & \raisebox{-0.5mm}{\boxed{\color{white}\rule{4.5mm}{4.5mm}}}
      & \raisebox{-1.8mm}{\rule{7mm}{7mm}}
        {\color{white}\mathclap{\mspace{-28mu} 1}}
      & \raisebox{-0.5mm}{\boxed{\color{white}\rule{4.5mm}{4.5mm}}}
      & \raisebox{-1.8mm}{\rule{7mm}{7mm}}
        {\color{white}\mathclap{\mspace{-28mu} -1}}
      &&& \vline\\
      &\rule[-2mm]{0mm}{9mm}
      & \raisebox{-1.8mm}{\rule{7mm}{7mm}}
        {\color{white}\mathclap{\mspace{-28mu} -1}}
      & \raisebox{-0.5mm}{\boxed{\color{white}\rule{4.5mm}{4.5mm}}}
      & \raisebox{-1.8mm}{\rule{7mm}{7mm}}
        {\color{white}\mathclap{\mspace{-28mu} 1}}
      & \raisebox{-0.5mm}{\boxed{\color{white}\rule{4.5mm}{4.5mm}}}
      &&& \vline\\
      &\rule[-2mm]{0mm}{9mm}
      & \raisebox{-0.5mm}{\boxed{\color{white}\rule{4.7mm}{4.5mm}}}
      & \raisebox{-1.8mm}{\rule{7mm}{7mm}}
        {\color{white}\mathclap{\mspace{-28mu} -1}}
      & \raisebox{-0.5mm}{\boxed{\color{white}\rule{4.5mm}{4.5mm}}}
      & \raisebox{-1.8mm}{\rule{7mm}{7mm}}
        {\color{white}\mathclap{\mspace{-28mu} 1}}
      &&& \vline\\
      &\rule[-2mm]{0mm}{9mm}
      & \raisebox{-1.8mm}{\rule{7mm}{7mm}}
        {\color{white}\mathclap{\mspace{-28mu} 1}}
      & \raisebox{-0.5mm}{\boxed{\color{white}\rule{4.5mm}{4.5mm}}}
      & \raisebox{-1.8mm}{\rule{7mm}{7mm}}
        {\color{white}\mathclap{\mspace{-28mu} -1}}
      & \raisebox{-0.5mm}{\boxed{\color{white}\rule{4.5mm}{4.5mm}}}
      &&&\vline\\
      &&&&&&\ddots&&\vline\\
      &&&&&&&&\vline&&\rule[-0.5em]{0.1mm}{0.5em}\\\cline{1-11}
      &&&&&&&&\vline&\phantom{\rule{5mm}{10mm}}&\vline\\\cline{9-12}
      &&&&&&&&\mspace{-0mu}\rule[8mm]{0.1mm}{0.5em}&
      & \vline&\phantom{\rule{9mm}{10mm}}\\
    \end{pmatrix}.
  \end{equation*}
  \caption{If the $n$'th \csproblem{1-in-3SAT} clause involves variables
    $i,j,k$, the construction encodes the coefficients from the
    inequalities of \cref{eq:clause_constraints} into the $n$'th
    on-diagonal $4\times 4$ block of $B^i$, $B^j$ and $B^k$. All other
    $B^c$ corresponding to variables that do not appear in that clause
    will have zeros in that particular block.}
  \label{fig:Bc_clause_block}

  \begin{equation*}
    B^c =
    \begin{pmatrix}
      \phantom{\rule{10mm}{10mm}}&\vline&&&&&&&&
      &\rule[-0.5em]{0.1mm}{0.5em}\\\cline{1-11}
      &\vline&&&&&&&&&\vline\\
      &\vline&&\ddots&&&&&&&\vline\\
      &\vline&&
      & \raisebox{-0.5mm}{\boxed{\color{white}\rule{4.5mm}{4.5mm}}}
      & \raisebox{-1.8mm}{\rule{7mm}{7mm}}
          {\color{white}\mathclap{\mspace{-28mu} 1}}
      & \raisebox{-0.5mm}{\boxed{\color{white}\rule{4.5mm}{4.5mm}}}
      & \raisebox{-1.8mm}{\rule{7mm}{7mm}}
          {\color{white}\mathclap{\mspace{-28mu} -1}}
      & \rule[-2mm]{0mm}{9mm}&&\vline\\
      &\vline&&
      & \raisebox{-1.8mm}{\rule{7mm}{7mm}}
          {\color{white}\mathclap{\mspace{-28mu} -1}}
      & \raisebox{-0.5mm}{\boxed{\color{white}\rule{4.5mm}{4.5mm}}}
      & \raisebox{-1.8mm}{\rule{7mm}{7mm}}
          {\color{white}\mathclap{\mspace{-28mu} 1}}
      & \raisebox{-0.5mm}{\boxed{\color{white}\rule{4.5mm}{4.5mm}}}
      & \rule[-2mm]{0mm}{9mm}&&\vline\\
      &\vline&&
      & \raisebox{-0.5mm}{\boxed{\color{white}\rule{4.5mm}{4.5mm}}}
      & \raisebox{-1.8mm}{\rule{7mm}{7mm}}
          {\color{white}\mathclap{\mspace{-28mu} -1}}
      & \raisebox{-0.5mm}{\boxed{\color{white}\rule{4.5mm}{4.5mm}}}
      & \raisebox{-1.8mm}{\rule{7mm}{7mm}}
          {\color{white}\mathclap{\mspace{-28mu} 1}}
      & \rule[-2mm]{0mm}{9mm}&&\vline\\
      &\vline&&
      & \raisebox{-1.8mm}{\rule{7mm}{7mm}}
          {\color{white}\mathclap{\mspace{-28mu} 1}}
      & \raisebox{-0.5mm}{\boxed{\color{white}\rule{4.5mm}{4.5mm}}}
      & \raisebox{-1.8mm}{\rule{7mm}{7mm}}
          {\color{white}\mathclap{\mspace{-28mu} -1}}
      & \raisebox{-0.5mm}{\boxed{\color{white}\rule{4.5mm}{4.5mm}}}
      &\rule[-2mm]{0mm}{9mm}&&\vline\\
      &\vline&&&&&&&\ddots&&\vline\\
      &\vline&&&&&&&&&\vline\\\cline{2-12}
      &\rule[9mm]{0.1mm}{0.5em}&&&&&&&&
      & \vline&\phantom{\rule{10mm}{10mm}}\\
    \end{pmatrix}.
  \end{equation*}
  \caption{Each $B^c$ contains a unique block of non-zero entries in the
    second set of on-diagonal $4\times 4$ blocks, corresponding to the
    \csproblem{1-in-3SAT} boolean constraints of
    \cref{eq:boolean_constraints}.}
  \label{fig:Bc_boolean_blocks}
\end{figure}

Colouring $Q$ in the same chess-board pattern, the contribution to its
``black squares'' from the first term of \cref{eq:special_case_Q} is
generated by the off-diagonal elements $\lambda_r$:
\begin{equation}
  \label{eq:symmetric_matrix}
  \sum_r\xx_r^{\vphantom{T}}\xx_r^T\otimes
  \begin{pmatrix}1&1\\1&1\end{pmatrix}\otimes
  \begin{pmatrix}\cdot&\lambda_r\\\lambda_r&\cdot\end{pmatrix}
  = S\otimes
    \begin{pmatrix}1&1\\1&1\end{pmatrix}\otimes
    \begin{pmatrix}\cdot&1\\1&\cdot\end{pmatrix}.
\end{equation}
(The dots emphasise that the ``white squares'' generated by those entries
will be specified later.) Since $\{\xx_r\}$ and $\{\lambda_r\}$ can be
chosen freely, the first tensor factor in this expression is just the
eigenvalue decomposition of an arbitrary real, symmetric matrix $S$. If we
choose the first $n_C$ diagonal elements of $S$ to be $1/2$, and choose
the next $n_v$ diagonal elements of $S$ to be $5/6$, then it is
straightforward to verify that the equations in the ccp condition of
\cref{eq:special_case_ccp1} corresponding to the ``black squares'' in
on-diagonal $4\times 4$ blocks are exactly the \csproblem{1-in-3SAT}
inequalities of \cref{eq:boolean_constraints,eq:clause_constraints} (see
\cref{fig:Q_clause_blocks,fig:Q_boolean_blocks}). Note that the
off-diagonal elements of $S$ are not specified yet.
\begin{figure}[!htbp]
  \centering
  \color{black}
  \begin{equation*}
    Q =
    \begin{pmatrix}
      \\
      &\ddots&&&&&&&\vline\\
      &\rule[-2mm]{0mm}{9mm}
      & \raisebox{-0.9mm}{\boxed{\color{white}\rule{4.5mm}{4.5mm}}}
      & \raisebox{-2.2mm}{\rule{7mm}{7mm}}
        {\color{white}\mathclap{\mspace{-28mu} -\frac{1}{2}}}
      & \raisebox{-0.9mm}{\boxed{\color{white}\rule{4.5mm}{4.5mm}}}
      & \raisebox{-2.2mm}{\rule{7mm}{7mm}}
        {\color{white}\mathclap{\mspace{-28mu} \frac{3}{2}}}
      &&& \vline\\
      &\rule[-2mm]{0mm}{9mm}
      & \raisebox{-2.2mm}{\rule{7mm}{7mm}}
        {\color{white}\mathclap{\mspace{-28mu} \frac{3}{2}}}
      & \raisebox{-0.9mm}{\boxed{\color{white}\rule{4.5mm}{4.5mm}}}
      & \raisebox{-2.2mm}{\rule{7mm}{7mm}}
        {\color{white}\mathclap{\mspace{-28mu} -\frac{1}{2}}}
      & \raisebox{-0.9mm}{\boxed{\color{white}\rule{4.5mm}{4.5mm}}}
      &&& \vline\\
      &\rule[-2mm]{0mm}{9mm}
      & \raisebox{-0.9mm}{\boxed{\color{white}\rule{4.5mm}{4.5mm}}}
      & \raisebox{-2.2mm}{\rule{7mm}{7mm}}
        {\color{white}\mathclap{\mspace{-28mu} \frac{3}{2}}}
      & \raisebox{-0.9mm}{\boxed{\color{white}\rule{4.5mm}{4.5mm}}}
      & \raisebox{-2.2mm}{\rule{7mm}{7mm}}
        {\color{white}\mathclap{\mspace{-28mu} -\frac{1}{2}}}
      &&& \vline\\
      &\rule[-2mm]{0mm}{9mm}
      & \raisebox{-2.2mm}{\rule{7mm}{7mm}}
        {\color{white}\mathclap{\mspace{-25mu} -\frac{1}{2}}}
      & \raisebox{-0.9mm}{\boxed{\color{white}\rule{4.5mm}{4.5mm}}}
      & \raisebox{-2.2mm}{\rule{7mm}{7mm}}
        {\color{white}\mathclap{\mspace{-25mu} \frac{3}{2}}}
      & \raisebox{-0.9mm}{\boxed{\color{white}\rule{4.5mm}{4.5mm}}}
      &&&\vline\\
      &&&&&&\ddots&&\vline\\
      &&&&&&&&\vline&&\rule[-0.5em]{0.1mm}{0.5em}\\\cline{1-11}
      &&&&&&&&\vline&\phantom{\rule{5mm}{10mm}}&\vline\\\cline{9-12}
      &&&&&&&&\mspace{-0mu}\rule[8mm]{0.1mm}{0.5em}&
      & \vline&\phantom{\rule{7mm}{10mm}}\\
    \end{pmatrix}.
  \end{equation*}
  \caption{The first set of on-diagonal $4\times 4$ blocks of $Q$ contain
    the constants for the \csproblem{1-in-3SAT} clause inequalities of
    \cref{eq:clause_constraints}\dots}
  \label{fig:Q_clause_blocks}

  \begin{equation}
    Q =
    \begin{pmatrix}
      \phantom{\rule{10mm}{10mm}}&\vline&&&&&&&&
      &\rule[-0.5em]{0.1mm}{0.5em}\\\cline{1-11}
      &\vline&&&&&&&&&\vline\\
      &\vline&&\ddots&&&&&&&\vline\\
      &\vline&&
      & \raisebox{-0.9mm}{\boxed{\color{white}\rule{4.5mm}{4.5mm}}}
      & \raisebox{-2.2mm}{\rule{7mm}{7mm}}
          {\color{white}\mathclap{\mspace{-28mu} \frac{1}{2}}}
      & \raisebox{-0.9mm}{\boxed{\color{white}\rule{4.5mm}{4.5mm}}}
      & \raisebox{-2.2mm}{\rule{7mm}{7mm}}
          {\color{white}\mathclap{\mspace{-28mu} \frac{7}{6}}}
      & \rule[-2mm]{0mm}{9mm}&&\vline\\
      &\vline&&
      & \raisebox{-2.2mm}{\rule{7mm}{7mm}}
          {\color{white}\mathclap{\mspace{-28mu} \frac{7}{6}}}
      & \raisebox{-0.9mm}{\boxed{\color{white}\rule{4.5mm}{4.5mm}}}
      & \raisebox{-2.2mm}{\rule{7mm}{7mm}}
          {\color{white}\mathclap{\mspace{-28mu} \frac{1}{2}}}
      & \raisebox{-0.9mm}{\boxed{\color{white}\rule{4.5mm}{4.5mm}}}
      & \rule[-2mm]{0mm}{9mm}&&\vline\\
      &\vline&&
      & \raisebox{-0.9mm}{\boxed{\color{white}\rule{4.5mm}{4.5mm}}}
      & \raisebox{-2.2mm}{\rule{7mm}{7mm}}
          {\color{white}\mathclap{\mspace{-28mu} \frac{7}{6}}}
      & \raisebox{-0.9mm}{\boxed{\color{white}\rule{4.5mm}{4.5mm}}}
      & \raisebox{-2.2mm}{\rule{7mm}{7mm}}
          {\color{white}\mathclap{\mspace{-28mu} \frac{1}{2}}}
      & \rule[-2mm]{0mm}{9mm}&&\vline\\
      &\vline&&
      & \raisebox{-2.2mm}{\rule{7mm}{7mm}}
          {\color{white}\mathclap{\mspace{-28mu} \frac{1}{2}}}
      & \raisebox{-0.9mm}{\boxed{\color{white}\rule{4.5mm}{4.5mm}}}
      & \raisebox{-2.2mm}{\rule{7mm}{7mm}}
          {\color{white}\mathclap{\mspace{-28mu} \frac{7}{6}}}
      & \raisebox{-0.9mm}{\boxed{\color{white}\rule{4.5mm}{4.5mm}}}
      &\rule[-2mm]{0mm}{9mm}&&\vline\\
      &\vline&&&&&&&\ddots&&\vline\\
      &\vline&&&&&&&&&\vline\\\cline{2-12}
      &\rule[9mm]{0.1mm}{0.5em}&&&&&&&&
      & \vline&\phantom{\rule{10mm}{10mm}}\\
    \end{pmatrix}.
  \end{equation}
  \caption{\dots whilst the second set of on-diagonal $4\times 4$ blocks
    of $Q$ contain the constants for the \csproblem{1-in-3SAT} boolean
    inequalities of \cref{eq:boolean_constraints}.}
  \label{fig:Q_boolean_blocks}
\end{figure}

We have successfully encoded the correct coefficients and constants into
certain matrix elements of $B^c$ and $Q$. But all the other elements of
these matrices also generate inequalities via
\cref{eq:special_case_ccp1}. To ``filter out'' these unwanted
inequalities, we choose the remaining diagonal elements and all
off-diagonal elements of the symmetric matrix $S$ to be large and
positive, thereby ensuring all unwanted inequalities are slack.

The matrices $A_c$ from \cref{eq:special_case_Ac} automatically satisfy
the normalisation condition of \cref{lem:L0+Ac_properties}, but $L_0$, as
constructed so far, will not. We use the ``white squares'' of $Q$ (see
\cref{fig:Q_clause_blocks,fig:Q_boolean_blocks}), generated by the
diagonal elements in the third tensor factors of
\cref{eq:special_case_Q}, to renormalise the column sums to zero. Recall
that both $\{\xx_r\}$ and $\{\vv_c,\vv_{c'}\}$ are complete sets of
mutually orthogonal vectors. Rearranging \cref{eq:special_case_Q}, $Q$ is
therefore given by
\begin{equation}\label{eq:constructed_N}
  Q = k\id + S\otimes\begin{pmatrix}1&1\\1&1\end{pmatrix}
             \otimes\begin{pmatrix}1&1\\1&1\end{pmatrix}
      + \sum_c\vv_c^{\vphantom{T}}\vv_c^T
        \otimes\begin{pmatrix}
          \phantom{-}1&-1\\-1&\phantom{-}1
        \end{pmatrix}
        \otimes\begin{pmatrix}
          0&-\frac{1}{3}\\\frac{1}{3}&\phantom{-}0
        \end{pmatrix}.
\end{equation}
Now, the only requirement on the off-diagonal elements of $S$ is that
they be sufficiently positive. Also, from the form of
\cref{eq:constructed_N}, the columns in any individual $4\times 4$ block
of $Q$ sum to the same value. Thus, by adjusting the elements of $S$, we
can ensure that all columns of $Q-k\id$ sum to the \emph{same} positive
value, which we call $\sigma$. Choosing $k=-\sigma$, the negative
on-diagonal element in each column (generated by the $k\id$ term) will
cancel the positive contribution from the off-diagonal elements, thereby
satisfying the normalisation condition, as required.

Finally, we must ensure that the second ccp condition of
\cref{eq:special_case_ccp2} is always satisfied, for which we require a
simple lemma.
\begin{lemma}
  \label{lem:Q+P}
  If $D\geq -\sigma\id$ is a diagonal $d\times d$-dimensional matrix, then
  there exists a symmetric matrix $P$ such that $ P_{i,i} = 0$ for all
  $i$ and
  \begin{equation}
    (\id-\vec{w}\vec{w}^T)(D+P)(\id-\vec{w}\vec{w}^T)\geq 0,
  \end{equation}
  where $\vec{w} = (1,1,\dots,1)^T/\sqrt{d}$.
\end{lemma}
\begin{proof}
  Choose $P = \alpha(\id-\vec{w}\vec{w}^T) +
  \alpha(1-d)\vec{w}\vec{w}^T$. Then the diagonal elements of $P$ are
  \begin{equation}
    P_{i,i}
    = \alpha\left(1-\frac{1}{d}\right) + \alpha(1-d)\frac{1}{d}
    = 0,
  \end{equation}
  and
  \begin{equation}
    (\id-\vec{w}\vec{w}^T)(D+P)(\id-\vec{w}\vec{w}^T)
    \geq  (\alpha-\sigma )(\id-\vec{w}\vec{w}^T),
  \end{equation}
  which is positive semi-definite for $\alpha\geq \sigma$.
\end{proof}

The coefficients $P_{i,j}$ in \cref{eq:special_case_L0} can be chosen
freely, since these coefficients play no role in either the normalisation
or in encoding \csproblem{1-in-3SAT}, so the matrix $P$ in the ccp
condition of \cref{eq:special_case_ccp2} can be chosen to be any matrix
with zeros down the main diagonal. \cref{eq:special_case_ccp2} is exactly
of the form given in \cref{lem:Q+P} with
\begin{equation}
  D_{i,i} = Q_{i,i}
\end{equation}
and choosing $P$ accordingly ensures that it is always satisfied.

\subsection{Perturbations}
In the discussion preceding the definition of \csproblem{Lindblad
  generator}, we argued that we need only consider non-singular,
non-degenerate channels. Generators of such channels are necessarily
bounded and non-degenerate as well, and the proof of equivalence of
\csproblem{Lindblad generator} and \csproblem{Markovian map}, leading to
\cref{thm:Lindblad-Markov_reduction}, breaks down if these properties do
not hold, since additional branches of the matrix logarithm arise:
applying an arbitrary similarity transformation to a degenerate Jordan
block will give another logarithm. The matrix $L_0$ we have constructed
is clearly bounded, but it is highly degenerate.

We will now slightly modify the above construction, removing the
mentioned degeneracies. In fact, most of the degeneracies can easily be
lifted by as large a margin as desired by perturbing suitable elements of
$L_0$, without affecting the conditions of
\cref{lem:L0+Ac_properties}. The only ones that require more care are
degeneracies due to the final two terms of \cref{eq:special_case_Q}, as
some of those matrix elements were used to encode \csproblem{1-in-3SAT}.

It is not difficult to verify that $m_c$ will be constrained to the same
set of integer values if the perturbation to any constant in the set of
inequalities is less than $1/6$ (the second inequality in
\cref{eq:boolean_constraints} being the most sensitive). The constants
are given directly by matrix elements of $L_0$, so we are free to lift
the remaining degeneracies in $L_0$ by perturbing each summand in the
final two terms of \cref{eq:special_case_Q} by a different amount, as
long as we ensure that no element of $L_0$ is perturbed by more than
$1/6$. This can be achieved by perturbing each off-diagonal
element\footnote{We avoid perturbing the diagonal elements, as that would
  make satisfying the normalisation condition far more difficult.} of the
final tensor factor by a different integer multiple of
\begin{equation}
  \frac{2}{9d}\begin{pmatrix}0&-1\\1&\phantom{-}0\end{pmatrix}.
\end{equation}
No element of $L_0$ is then perturbed by more than $1/18$ (this is
deliberately stricter than necessary by a factor of three, for reasons
that will become clearer later), and the minimum eigenvalue separation
for the perturbed $L_0$ is $2/(9d)$.

By construction, $L_0$ is a Lindblad generator iff the original
\csproblem{1-in-3SAT} instance was satisfiable, so we have achieved the
first half of the reduction. It remains to choose a value of $\delta$
such that this also holds for any $L'_0$ in the $\delta$-ball around
$L_0$. As noted above, the inequalities in
\cref{eq:boolean_constraints,eq:clause_constraints} are insensitive to
small perturbations. Specifically, one can verify that the set of
feasible $m_c$ will be unchanged if each coefficient and constant (this
time including zero coefficients, i.e.\ coefficients of variables that do
not appear explicitly in
\cref{eq:boolean_constraints,eq:clause_constraints}) is perturbed by less
than $\min[1/18(n_v+1),5/18(2n_v+1)]$. (Recall that we already perturbed
the constants by (up to) $1/18$ to lift eigenvalue degeneracies. This
bound is deliberately stronger by a factor of two than would appear to be
necessary at this stage, but in any case it is certainly stronger than is
strictly necessary.)

The constants in the inequalities are given by matrix elements of
$L_0$. If we choose the norm in \csproblem{Lindblad generator} to be the
$l_\infty$ norm, then is is sufficient to require
\begin{equation}\label{eq:delta_bound_L0}
  \delta \leq
  \min\left[\frac{1}{18(n_v+1)},\frac{5}{18(2n_v+1)}\right].
\end{equation}
The coefficients in the inequalities are given by matrix elements of
$A_c$, which are formed from the eigenvectors of $L_0$. Thus, to bound
perturbations of the coefficients, we must bound perturbations of the
eigenvectors in terms of the perturbation to $L_0$, which is less
trivial. We will need the following result from Ref.~\cite{Stewart}, and
a simple corollary.
\begin{lemma}
  \label{lem:eigenspace_perturbation}
  Suppose $A$ is a normal matrix, with $E$ an arbitrary matrix of the
  same dimension. Let $Q=(\vv_1,Q_2)$ be unitary, such that $\vv_1$ is
  an eigenvector of $A$, and partition the matrix $Q^\dagger EQ$
  conformally with $Q^\dagger AQ$, so that\footnote{$Q^\dagger AQ$
    must be of this form, as the Schur decomposition of a normal
    matrix is diagonal.}:
  \begin{equation}
    Q^\dagger AQ = \begin{pmatrix}\lambda_1 &0\\0&A_{2,2}\end{pmatrix},
    \quad
    Q^\dagger EQ = \begin{pmatrix}E_{1,1}&E_{1,2}\\
                                E_{2,1}&E_{2,2}\end{pmatrix},
  \end{equation}
  where $\{\lambda_i\}$ denote the eigenvalues of $A$, with $\lambda_1$
  the eigenvalue associated with $\vv_1$. Let
  \begin{equation}\label{eq:Delta}
    \Delta = \min_{i\neq 1}\Abs{\lambda_1 - \lambda_i}
             - \MatnormF{E_{1,1}} - \MatnormF{E_{2,2}},
  \end{equation}
  where $\matnormF{X}^2= \sum_{i,j}\abs{X_{i,j}}^2$ is the
  Frobenius (or Hilbert-Schmidt) norm. If $\Delta > 0$, and
  \begin{equation}\label{eq:eigenspace_perturbation_requirement}
    \frac{\MatnormF{E_{2,1}}\MatnormF{E_{1,2}}}{\Delta^2}
    \leq \frac{1}{4},
  \end{equation}
  then there exists a matrix $P$ satisfying
  \begin{equation}
    \MatnormF{P}\leq 2\frac{\MatnormF{E_{2,1}}}{\Delta}
  \end{equation}
  such that $\vv' = (\vv_1+Q_2P)(\id+P^\dagger P)^{-1/2}$
  is a unit eigenvector of $A+E$ (in the Frobenius norm).
\end{lemma}
\begin{proof}
  This is a slight generalisation of Theorem~8.1.12 from
  Ref.~\cite{Golub+vanLoan}, or slight restriction of Theorem~4.11
  from Ref.~\cite{Stewart}, to the case of normal $A$.
\end{proof}

\begin{corollary}
  \label{cor:eigenprojector_perturbation}
  Suppose $A$ is a normal matrix, with $E$ an arbitrary matrix of the
  same dimension. If $\vv$ is a unit (in Frobenius norm) eigenvector
  of $A$ associated with a non-degenerate eigenvalue, and the
  requirements of \cref{lem:eigenspace_perturbation} are fulfilled,
  then there exists a unit eigenvector $\vv'$ of $A+E$ such that
  \begin{gather}
    \MatnormF{\vv\vv^\dagger-\vv'\vv'^\dagger} \leq K\MatnormF{E},
  \end{gather}
  with
  \begin{equation}
     K = \frac{4\left(d\MatnormF{E}+\sqrt{d-1}\Delta\right)}
              {\Delta^2-4\MatnormF{E}^2}
  \end{equation}
  and $\Delta$ as defined in \cref{lem:eigenspace_perturbation}.
\end{corollary}
\begin{proof}
  From \cref{lem:eigenspace_perturbation}, we have
  \begin{align}
    \MatnormF{\vv'\vv'^\dagger - \vv\vv^\dagger}
    &= \MatnormF{\frac{(\vv_1+Q_2P)(\vv_1+Q_2P)^\dagger}{\id+P^\dagger P}
                -\vv_1\vv_1^\dagger}\\
    &\leq \frac{2\matnormF{\vv}\matnormF{Q_2}
              + \matnormF{P}\left(
                \matnormF{\vv}^2 + \matnormF{Q_2}^2\right)}
             {1 - \matnormF{P^\dagger P} \rule{0em}{1em}}
           \MatnormF{P}\\
    &\leq \frac{2\sqrt{d-1} + d\matnormF{P}}
            {1-\matnormF{P}^2}
       \matnormF{P}.
  \end{align}
  in which we have used Lemma~2.3.3 from Ref.~\cite{Golub+vanLoan} to
  bound $(\id+P^\dagger P)^{-1}$, and the fact that
  $\matnormF{U}=\sqrt{d}$ for any $d\times d$ unitary $U$. The result
  follows by substituting the bound on $\matnormF{P}$ from
  \cref{lem:eigenspace_perturbation}, and using
  $\matnormF{E_{2,1}}\leq\matnormF{E}$.
\end{proof}

Now, each $A_c$ is a sum of two eigenprojectors, and $L_0$ happens to be
normal. Applying \cref{cor:eigenprojector_perturbation}, and using the
fact that $\matnorm{X}_\infty \leq \matnormF{X}$, we see that it suffices
to restrict
\begin{equation}\label{eq:delta_bound_Ac}
  \delta \leq \frac{1}{2K}
    \min\left[\frac{1}{18(n_v+1)},\frac{5}{18(2n_v+1)}\right].
\end{equation}
We must also satisfy the two requirements of
\cref{lem:eigenspace_perturbation}. Recalling that the minimum eigenvalue
separation of $L_0$ is $2/(9d)$, we see that it is sufficient to impose
\begin{equation}
  \delta < \frac{1}{9d^2}
  \quad\text{and}\quad
  \delta \leq \frac{\min_{i\neq j}\Abs{\lambda_i-\lambda_j}}{4d}
          = \frac{1}{18d}\label{eq:delta_bound_lemma}.
\end{equation}

For $L_0$, satisfying the inequalities is equivalent to satisfying the
ccp condition of \cref{lem:Lindblad}. However, even choosing $\delta$ to
satisfy \cref{eq:delta_bound_L0,eq:delta_bound_Ac,eq:delta_bound_lemma},
this may no longer be the case for all $L'_0$ within the $\delta$-ball
around $L_0$. If the inequalities are infeasible, then at least one
diagonal element of any $(\id-\omega)L'_m{\!}^\Gamma(\id-\omega)$ must be
negative, and it is still the case that the ccp condition is violated
(since non-negativity of the diagonal elements is a necessary condition
for a matrix to be positive semi-definite). But if the inequalities
\emph{can} be satisfied, the most we can say is that all diagonal
elements of $(\id-\omega)L'_m{\!}^\Gamma(\id-\omega)$ are lower-bounded by
$1/18$.

Now
\begin{equation}
  L'_m = L'_0 +\sum_cm_cA'_c
\end{equation}
with $0 \leq m_c \leq 1$ integer, and the $A'_c$ are perturbations of
$A_c$. The off-diagonal elements of the latter are zero. Therefore, we
can control the magnitude of the off-diagonal elements of the $n_v$
different $A'_c$ by applying \cref{cor:eigenprojector_perturbation}
again, whilst controlling the off-diagonal elements of $L'_0$ by
restricting $\delta$ directly, as before. Putting all this together, we
see that imposing
\begin{equation}\label{eq:delta_bound_diag_dom}
    \delta \leq \frac{1}{18d}\quad\text{and}\quad
    \delta \leq \frac{1}{32K n_v d}
\end{equation}
ensures that the off-diagonal elements of any $L'_m$ are upper-bounded by
$1/(18d)$. However, this implies that
$(\id-\omega)L'_m{\!}^\Gamma(\id-\omega)$ is diagonally-dominant, which is
sufficient to guarantee positive-semi-definiteness.

Thus, if $\delta>0$ is chosen to satisfy
\cref{eq:delta_bound_L0,eq:delta_bound_Ac,eq:delta_bound_lemma,eq:delta_bound_diag_dom},
then for any $L'_0$ within a $\delta$-ball around $L_0$ (in the
$l_\infty$ norm), satisfying the ccp condition is equivalent to
satisfying the original \csproblem{1-in-3SAT} problem. Comparing the
bounds on $\delta$ from
\cref{eq:delta_bound_L0,eq:delta_bound_Ac,eq:delta_bound_lemma,eq:delta_bound_diag_dom},
we have
\begin{equation}
  \delta = \order{n_v^{-1}(n_C+2n_v)^{-3}}.
\end{equation}
Sufficient bounds for any other norm can easily be obtained via
equivalence of norms in finite-dimensional spaces, and will at worst
introduce additional factors polynomial in the dimension (i.e.\
polynomial in $n_v$ and $n_C$). The fact that $\delta^{-1}$ has to scale
only polynomially makes our results far more compelling; it cannot be
claimed that they are a consequence of unreasonable precision
demands. Even this mild scaling may be an artifact of the construction,
and it would be interesting to know if a construction exists in which
$\delta$ can be taken constant.

Finally, it remains to consider the promise required in the definition of
\csproblem{Lindblad generator}. Assume that the promise is \emph{not}
satisfied. In that case, $L_0$ itself clearly cannot be the generator of
a CPT map. But $L_0$ satisfies the Hermiticity and normalisation
requirements of \cref{lem:Lindblad} by construction, so it must fail to
satisfy the ccp condition. Thus failing to satisfy the promise implies
that the \csproblem{1-in-3SAT} instance must have been
unsatisfiable. Combining the arguments used in the proofs of
\cref{thm:Lindblad-Markov_reduction,thm:Markov_map-channel_equivalence}
gives an efficient procedure for deciding whether $(L_0,\delta)$
satisfies the promise, thereby deciding these instances. This leaves only
instances that do satisfy the promise, as required.

We have reduced satisfiable instances of \csproblem{1-in-3SAT} to
\csproblem{Lindblad generator} instances that return the first assertion,
and have either efficiently decided unsatisfiable instances of
\csproblem{1-in-3SAT} (because they fail to satisfy the
promise)\footnote{It is amusing, but probably of no practical value, to
  note that this provides a new ``gadget'' for efficiently deciding
  certain non-satisfiable instances of \csproblem{1-in-3SAT}.}, or
reduced them to \csproblem{Lindblad generator} instances that return the
second assertion. This completes the proof that
\begin{lemma}
  \csproblem{1-in-3SAT} $\leq$ \csproblem{Lindblad generator}
\end{lemma}
and, since \csproblem{1-in-3SAT} is NP-complete,
\begin{corollary}\label{cor:Lindblad_NP-hardness}
  \csproblem{Lindblad generator} is NP-hard.
\end{corollary}
But, by the chain of equivalences proven in
\cref{thm:Markov_map-channel_equivalence,cor:Lindblad-Markov_equivalence},
this implies our main result:
\begin{theorem}\label{thm:Markov_NP-hardness}
  \csproblem{Markovian channel} and \csproblem{Markovian map} are NP-hard.
\end{theorem}

\Cref{thm:Markov_NP-hardness} tells us that the Markovianity problem is
NP-hard. What of the more general question of determining whether a given
family of maps are members of the same continuous, one-parameter,
completely positive semi-group? Formulated rigorously, this is a
generalised version of \csproblem{Markovian map}, in which a family of
maps $E_t$ is given, along with their associated times $t$ (up to some
precision), and the answer should assert the existence or otherwise of a
\emph{common} Lindblad generator for all the maps up to precision
$\varepsilon>0$ (or assert that at least one of the $E_t$ is not CPT up
to precision $\varepsilon'>0$).

A first trivial observation is that, since we know there exists a special
case of this problem that is NP-hard, namely \csproblem{Markovian map}
itself, the general problem is automatically NP-hard. However, this
leaves open the question of whether the complexity depends on the number
of maps in the family. Recalling the physical motivation behind the
problem, one might expect that, given more information about the dynamics
(e.g.\ by taking many tomographic snapshots), the problem would become
easier to resolve.

In fact, in proving the NP-hardness of \csproblem{Markovian map}, we have
already done all the work necessary to prove NP-hardness of the general
problem for any number of maps. Instead of computing a single map $E=e^L$ to
reduce \csproblem{Lindblad generator} to \csproblem{Markovian map}, we can
compute a family of any number of maps $E_t=e^{Lt}$. (To make this rigorous,
the arguments of \cref{thm:Lindblad-Markov_reduction} can straightforwardly be
extended to the case of a family of maps $E_t$.) So the problem for an
arbitrary (finite) number of maps is essentially no different to the problem
for a single map as far as the worst-case complexity is concerned.

\section{An Algorithm}
\label{sec:fixed_dimension}

The NP-hardness proof of \cref{sec:NP-hardness} implies that we are
unlikely to find an efficient algorithm for solving the Markovianity
problem. Nonetheless, there are two reasons to develop an algorithm for
solving it, even though it will be inefficient. The first reason is in
some sense a technicality. We would like to prove that solving the
Markovianity problem is equivalent to solving P=NP\@. That is, we want to
show that (i)~any efficient algorithm for solving the Markovianity
problem would imply P=NP, and conversely (ii)~\emph{if P=NP then there
  exists an efficient algorithm for solving the Markovianity problem}.
NP-hardness proves (i). But the weak-membership formulations of the
Markovianity problem (\csproblem{Markovian Channel/Map}) are not
technically members of the class NP, thus it is not clear whether proving
P=NP would be sufficient to provide an efficient algorithm for solving
them. Weak-membership problems do not belong to NP, for the simple reason
that NP is a decision class, but weak-membership problems are not
decision problems since they have instances in which both ``yes'' and
``no'' answers are simultaneously valid. (As mentioned above, the
appropriate complexity class for weak-membership problems is called
promise-NP; the additional promise is that the instance will not be one
of the ambiguous ones.) Giving an explicit algorithm for
\csproblem{Markovian Channel} which reduces to solving an NP-complete
problem resolves this technicality.

The second reason for developing an algorithm is that the NP-hardness
proof of \cref{sec:NP-hardness} requires the dimension to scale
polynomially with the size of the \csproblem{1-in-3SAT} problem being
encoded. So, although the general Markovianity problem for CPT maps and
embedding problem for stochastic matrices are NP-hard, it is interesting
to ask how the complexity scales if the dimension is fixed (in which case
the problem size scales only with the precision). By giving an explicit
algorithm, we show that \emph{for fixed dimension} the Markovianity
problem can be solved efficiently, i.e.\ the complexity scales only
polynomially with the precision. This is also the basis for the proposed
measure of Markovianity in Ref.~\cite{markovianity}.

One motivation for considering the case of fixed dimension is current
experimental limitations. A snapshot of a quantum evolution is measured
by performing full quantum process tomography. Tomography of a
$d$--dimensional system requires measuring a total of $d^4-d^2$ different
expectation values~\cite[\S 8.4.2]{Nielsen+Chuang}, and the expectation
value of each observable must be estimated by averaging over many runs.
The experimental overhead for all of this scales polynomially with the
dimension of the system, but a polynomial scaling can still be
prohibitive in practice! Current experiments can only perform full
process tomography for systems up to a few qubits, before the time
required becomes exorbitant. It is quite reasonable in this context to
regard dimension as a fixed parameter.

Since \csproblem{Markovian map} is equivalent to \csproblem{Markovian
  channel} by \cref{thm:Markov_map-channel_equivalence}, a
\csproblem{Markovian map} instance can be solved by first efficiently
reducing it to \csproblem{Markovian channel}, then solving the
\csproblem{Markovian channel} instance. We now describe an algorithm
which solves \csproblem{Markovian channel} in polynomial time for fixed
dimension. (The present treatment presents a detailed and rigorous proof
of the result already reported in Ref.~\cite{markovianity}.) It is not
difficult to adapt this algorithm to the classical
\csproblem{Embeddability} problem of \cref{sec:classical}. For
convenience, we will take the matrix norm in the definition of
\csproblem{Markovian channel} to be the Frobenius norm
$\matnormF{.}$.\footnote{It is straightforward to generalise these
  results to other norms.}

\begin{algorithm}[MARKOVIAN CHANNEL]
  \label{alg:Markov}
  \textbf{Input:} $(E,\varepsilon)$: Quantum channel $E$, precision
  $\varepsilon$.\\
  \textbf{Output:} One of the two assertions from
  \cref{prob:Markov_channel}.

  \addtolength{\parskip}{10em}
  \setlength{\abovedisplayskip}{0.3em}
  \setlength{\belowdisplayskip}{0.3em}
  \begin{algorithmic}[1]
    \State Calculate approximations $\bar{L}_0$ and $\bar{A}_c$ to $L_0 =
    \log E$ and $A_c$ (cf.\ \cref{lem:Lindblad}) to any precision
    $\kappa$, so that $\matnormF{\bar{L}_0-L_0} \leq \kappa$ and
    $\matnormF{\bar{A}_c-A_c} \leq \kappa$ ($\bar{L}_0$ and $\bar{A}_c$
    can be obtained e.g.\ by calculating the eigenvalues and eigenvectors
    of $E$).
    \label{line:bootstrap}
    %
    \State Calculate $\tdelta$ by solving
    \begin{equation}
      \begin{split}
        \exp\Bigl(\matnormF{\bar{L_0}}+M\sum_c\matnormF{\bar{A}_c}\Bigr)
          \exp\Bigl(\kappa+\frac{Md\kappa}{2}\Bigr)\,
          \tdelta\,e^{\tdelta}
         = \varepsilon,
      \end{split}\raisetag{1.5em}
    \end{equation}
    where $M$ depends polynomially on $\varepsilon$ (discussed in more
    detail below) and $d$ is the dimension of $E$.
    \label{line:delta}
    \medskip
    \State Calculate approximations $\tilde{\lambda}_i$ to the logarithms
    $\lambda_i$ of eigenvalues $e^{\lambda_i}$ of $E$, and to the
    eigenprojectors $\ketbra{\tilde{r_i}}{\tilde{l}_i}$ of $E$, to
    precision sufficient to ensure that
    \begin{gather}
      \biggMatnorm{
        \sum_i\tilde{\lambda}_i\ketbra{\tilde{r_i}}{\tilde{l}_i}
        - \sum_i\lambda_i\ketbra{r_i}{l_i}}
        \leq \frac{\tdelta}{12d\matnormF{\id-\omega}^3},\\
      \MatnormF{\ketbra{\tilde{r_i}}{\tilde{l}_i} - \ketbra{r_i}{l_i}}
        \leq \frac{\tdelta}{24\pi Md^2\matnormF{\id-\omega}^3},\\
      \abs{\tilde{\lambda}_i-\lambda_i}
        < \min_{j\neq k}\frac{\tilde{\lambda}_j-\tilde{\lambda}_k}{4}.
    \end{gather}
    \label{line:eigs}
    \medskip
    \State Use the results to calculate $\tL_0 =
    \sum_i\tilde{\lambda}_i\ketbra{\tilde{r_i}}{\tilde{l}_i}$ and the
    corresponding $\tA_c$ (cf.\ \cref{lem:Lindblad}).
    \label{line:L_approx}
    \medskip
    \State Solve the following mixed integer semi-definite program, in
    integer variables $m_c$ and real variable $t$:
    \setlength{\belowdisplayskip}{-0.3em}
    \begin{alignat*}{2}
      &\text{minimise}&\quad& t\\
      &\text{subject to}&\quad&
      (\id-\omega)\tL_0{\!}^\Gamma(\id-\omega)
      + \sum_cm_c(\id-\omega)\tA_c{\!}^\Gamma(\id-\omega)
      + t\id \geq 0.
    \end{alignat*}
    \label{line:integer_program}
    \If{$t \leq -\tdelta/(6d\matnormF{\id-\omega})$\label{line:if}}
      \State \Return ``Markovian'' (1\textsuperscript{st} assertion of
      \cref{prob:Markov_channel}).
    \ElsIf{$t > \tdelta/(6d\matnormF{\id-\omega})$}
      \State \Return ``non-Markovian'' (2\textsuperscript{nd} assertion
      of \cref{prob:Markov_channel}).
    \ElsIf{$t \leq \tdelta/(3d\matnormF{\id-\omega})$
           \label{line:boundary_case}}%
      \State \Return ``Markovian'' (1\textsuperscript{st} assertion of
      \cref{prob:Markov_channel}).\label{line:endif}
    \EndIf
  \end{algorithmic}
\end{algorithm}
\vspace{-8em} 

To prove correctness of \cref{alg:Markov}, first note that, from
lines~\ref{line:delta} to~\ref{line:L_approx},\linebreak
$\matnormF{\tL_0-L_0} \leq \tdelta/(12d\matnormF{\id-\omega}^3)$. Also, if
$\max_cm_c \leq M$, then from line~\ref{line:eigs} we have
\begin{equation}\label{eq:Lm_bound}
  \begin{split}
    \matnormF{\tL_m-L_m}
    &\leq \matnormF{\tL_0-L_0}
      +2\pi\sum_c\abs{m_c}\matnormF{\ketbra{\tilde{r_i}}{\tilde{l}_i}
                                   -\ketbra{r_i}{l_i}}\\
    &= \frac{\tdelta}{6d\matnormF{\id-\omega}^3}.
  \end{split}
\end{equation}
We will assume throughout the following that $M$ is an upper bound on the
values $m_c$ returned by the integer program of
line~\ref{line:integer_program}, i.e.\ that $\max_c\abs{m_c} \leq M <
\infty$, an assumption that will be justified later.

Now consider the three cases in lines~\ref{line:if}
to~\ref{line:endif}. To deal with the first two, we will need the
following simple lemma (see e.g.\ Ref.~\cite[Corollary
6.3.4]{Horn+Johnson}):
\begin{lemma}
  \label{lem:eigenvalue_perturbation}
  Let $A$ be normal, $E$ be an arbitrary matrix. If $\lambda'$ is an
  eigenvalue of $A+E$, then there exists some eigenvalue $\lambda$ of
  $A$ such that $\abs{\lambda'-\lambda}\leq\matnormF{E}$.
\end{lemma}
If $t \leq -\tdelta/(6d\matnormF{\id-\omega})$, then, from the definition
of the integer program in line~\ref{line:integer_program} of
\cref{alg:Markov}, we know that all eigenvalues of
$(\id-\omega)\tL_m^\Gamma(\id-\omega)$ are greater than
$\tdelta/(6d\matnormF{\id-\omega})$. Also, from \cref{eq:Lm_bound},
$\matnormF{(\id-\omega)(\tL_m^\Gamma-L_m^\Gamma)(\id-\omega)}
\leq\tdelta/(6d\matnormF{\id-\omega})$. \Cref{lem:eigenvalue_perturbation}
then implies that the minimum eigenvalue of
$(\id-\omega)L_m^\Gamma(\id-\omega)$ is non-negative, i.e.\ $L_m$ is
ccp. $L_0$ is therefore a Lindblad generator by \cref{lem:Lindblad}, thus
the original channel $E$ must itself be Markovian. Similarly, if $t >
\tdelta/(6d\matnormF{\id-\omega})$, then the minimum eigenvalue of
\emph{any} $(\id-\omega)L_m^\Gamma(\id-\omega)$ is strictly negative. Thus
all $L_m$ fail the ccp condition of \cref{lem:Lindblad}, $L_0$ is not a
Lindblad generator, and the original channel $E$ is non-Markovian.

Dealing with the final case in line~\ref{line:boundary_case} of
\cref{alg:Markov} requires the following result:
\begin{lemma}
  \label{lem:Lindblad_perturbation}
  If $L$ is Hermitian and normalised (in the sense of
  \cref{lem:Lindblad}), and the minimum eigenvalue of
  $(\id-\omega)L^\Gamma(\id-\omega)$ is bounded by $\lambda_{\mathrm{min}}
  \geq -\varepsilon$, then there exists a Lindblad generator $L'$ such that
  $\matnormF{L'-L} \leq \varepsilon\,d\,\matnormF{\id-\omega}$, where $d$ is
  the dimension of $L$.
\end{lemma}
\begin{proof}
  Consider the map $L' = L + \varepsilon(d\,\omega - d\id)$. Since $L$ is
  Hermitian and normalised in the above sense,
  we have $({L'}^\Gamma)^\dagger =
  {L'}^\Gamma$ and $\bra{\omega}L' = 0$, so these properties carry
  over to $L'$. But we also have
  \begin{equation}
    \begin{split}
      (\id-\omega){L'}^\Gamma(\id-\omega)
      &= (\id-\omega)L^\Gamma(\id-\omega)
         + \varepsilon(\id-\omega)(\id-d^2\omega)(\id-\omega)\\
      &= (\id-\omega)L^\Gamma(\id-\omega) + \varepsilon(\id-\omega).
    \end{split}
  \end{equation}
  Since $(\id-\omega)L^\Gamma(\id-\omega)$ has support only on the
  orthogonal complement of $\ket{\omega}$, and $(\id-\omega)$ acts as
  identity on that subspace, the minimum eigenvalue of
  $(\id-\omega){L'}^\Gamma(\id-\omega)$ is non-negative. Thus $L'$ also
  satisfies the ccp condition, and, by \cref{lem:Lindblad}, is a Lindblad
  generator.
\end{proof}

If $t \leq \tdelta/(3d\matnormF{\id-\omega})$, then the minimum eigenvalue
of $(\id-\omega)\tL_m^\Gamma(\id-\omega)$ is greater than
$-\tdelta/(3d\matnormF{\id-\omega})$, thus
\cref{lem:eigenvalue_perturbation,eq:Lm_bound} imply that the minimum
eigenvalue of $(\id-\omega)L_m^\Gamma(\id-\omega)$ is \emph{lower}-bounded
by
\begin{equation}
  \lambda_{\mathrm{min}} \geq -\tdelta/(3d\matnormF{\id-\omega})
    - \tdelta/(6d\matnormF{\id-\omega})
  = - \tdelta/(2d\matnormF{\id-\omega}).
\end{equation}
Applying \cref{lem:Lindblad_perturbation} to $L_m$ yields a Lindblad
generator $L'$ such that $\matnormF{L'-L_m} \leq
d\matnormF{\id-\omega}\tdelta/(d\matnormF{\id-\omega}) = \tdelta$ and,
since $L'$ is a Lindblad generator, $E'=e^{L'}$ is a Markovian
channel. But, using \cref{lem:exp_continuity}, we have
\begin{equation}
  \begin{split}
    \mspace{20mu}&\mspace{-20mu}\matnormF{E' - E}
    \leq e^{\matnormF{L_m}}e^{\matnormF{L'-L_m}}\matnormF{L'-L_m}\\
    &\leq \exp\Bigl(\matnormF{L_0}+M\sum_c\matnormF{A_c}\Bigr)
       \tdelta\,e^{\tdelta}\\
    &\leq \exp\biggl(\matnormF{\tL_0}+M\sum_c\matnormF{\tilde{A}_c}\biggr)
      \exp\left(\kappa+\frac{Md\kappa}{2}\right)\tdelta\,e^{\tdelta}\\
    &= \varepsilon,
  \end{split}\raisetag{7em}
\end{equation}
(with the inequality in the penultimate line resulting from
line~\ref{line:bootstrap} of \cref{alg:Markov}---recall that there are at
most $d/2$ matrices $\tilde{A}_c$---and the final equality from
line~\ref{line:delta}). Therefore, $E'$ is a Markovian channel within
distance $\varepsilon$ of the original channel $E$, and the first
assertion of \cref{prob:Markov_channel} is valid.

This proves correctness of \cref{alg:Markov}. What of its run-time? All
but a few steps can obviously be performed in polynomial-time. Recall
that we are assuming, without loss of generality, that $E$ is
non-degenerate and non-singular, which, more rigorously stated, requires
the condition number of $E$ to be upper-bounded by some constant. The
eigenvalue and eigenvector calculations of $E$ in lines~\ref{line:eigs}
and~\ref{line:bootstrap} can therefore be done efficiently in
$\varepsilon^{-1}$ and also the dimension~\cite[\S 7.2]{Golub+vanLoan}, with
the eigenvalue and eigenvector condition numbers of $E$~\cite[\S
7.2.2--5]{Golub+vanLoan} contributing a (possibly large) constant factor.

A question arises in calculating $\tilde{A}_c$: $\tL_0$ is not
necessarily a Hermitian map, so how can the eigenvalue pairs from which
to form $\tilde{A}_c$ (cf.\ \cref{eq:A_c}) be identified? But $L_0$
\emph{is} Hermitian, and the bound on $\abs{\tilde{\lambda}_i-\lambda_i}$
in line~\ref{line:eigs} ensures that the
$2\matnormF{\tilde{\lambda}_i-\lambda_i}$-disc around $\lambda_i^*$,
within which the conjugate partner of $\lambda_i$ must lie, is guaranteed
to contain a single $\tilde{\lambda}_j$, allowing approximately conjugate
pairs of eigenvalues to be identified.

The key step in the algorithm is the mixed integer semi-definite program
in line~\ref{line:integer_program}. (If \cref{alg:Markov} is adapted to
solve the classical \csproblem{Embeddability} problem, this becomes a
mixed \emph{linear} integer program instead.) In a generalisation of a
famous result by Lenstra~\cite{Schrijver} for linear integer programming,
Khachiyan and Porkolab proved that for any \emph{fixed} number of
variables, integer semi-definite feasibility problems can be solved in
polynomial time~\cite{Khachiyan+Porkolab,Porkolab}. In our case, fixing
the number of variables corresponds to fixing the system's dimension. The
integer semi-definite program can therefore be solved by applying the
Khachiyan-Porkolab algorithm to the feasibility problem for given $t$,
combined with binary search on $t$. From Corollary~1.3 of
Ref.~\cite{Khachiyan+Porkolab}, the run-time of the Khachiyan-Porkolab
part scales polynomially with the number of digits of precision to which
the elements of the coefficient matrices are specified. But the
coefficient matrices in our case are $\tL_0$ and $\tilde{A}_c$, and their
description size is independent of the precision to which the original
$E$ was specified, depending only on the precision parameter
$\varepsilon$. So the run-time of the Khachiyan-Porkolab step scales
polynomially in $\varepsilon^{-1}$, as required.

We can now also justify the assumption that an upper bound
$\max_c{m_c} \leq M$ can be placed on the integers $m_c$ resulting
from the integer program. Theorem~1.1 of
Ref.~\cite{Khachiyan+Porkolab} proves that such a bound exists and, in
the case of integer semi-definite programming
(\cite[Corollary~1.3]{Khachiyan+Porkolab}), that it scales as
\begin{equation}
  \log \max_c \abs{m_c} = 2^{\order{d^4}}\log l,
\end{equation}
where $l$ is the maximum bit-length of the entries of the coefficient
matrices $\tL_0$ and $\tilde{A}_c$, and we have translated other
parameters into our notation. Since we have already argued that the size
of the description of these matrices scales polynomially with
$\varepsilon^{-1}$, this gives a bound $M$ that scales as
\begin{equation}
  \max_c\abs{m_c} = \varepsilon^{(2^{\order{d^4}})\order{1}} = M,
\end{equation}
i.e.\ polynomially in $\varepsilon^{-1}$ as claimed.

Since the calculations in each line of \cref{alg:Markov} have run-times
that scale at most polynomially in $\varepsilon^{-1}$, and are independent
of the number of digits to which $E$ was specified, the entire algorithm
has run-time polynomial in the precision and independent of the size of
the description of $E$. This, together with
\cref{thm:Markov_map-channel_equivalence}, proves the main practical
result of this section:

\begin{theorem}
  For any fixed dimension, \csproblem{Markovian channel} and\linebreak
  \csproblem{Markovian map} can be solved in a run-time that scales
  polynomially in both the problem size (the size of the description of
  the channel) and the precision parameter $\varepsilon^{-1}$.
\end{theorem}

It is worth remembering that proving an algorithm has polynomial run-time
does not necessarily imply that it is the best algorithm to use in
practice. In fact, considering the first few branches of the logarithm is
often sufficient for practically relevant cases. Indeed, it would be
interesting to try to flesh out heuristics or a proof as to why this
simple approach is so successful. If $E$ is an experimentally measured
tomographic snapshot, the truncation errors in computing $\log E$, that
\cref{alg:Markov} expends much effort in accounting for, will, in all
likelihood, be swamped by experimental error. It is probably reasonable
to calculate $L_0$ and $A_c$ numerically, without worrying about
numerical errors, and solve the resulting mixed integer semi-definite
program using standard integer programming algorithms (which work well in
practise even though their scaling may theoretically not be polynomial in
the precision). If the $t$ thus obtained is comparable to the estimated
error, the most reasonable conclusion is that the experimental data
simply are not precise enough to give any definitive answer. In fact, a
more sophisticated answer is to quote the value of $t$ itself, as it is
(related to) a natural measure of ``Markovianity''. This is discussed in
more detail in Ref.~\cite{markovianity}.

All the steps of \cref{alg:Markov} also scale efficiently with the
dimension of $E$, apart from solving the mixed integer semi-definite
program in line~\ref{line:integer_program}. Since integer semi-definite
programming is in NP, this (together with
\cref{thm:Markov_map-channel_equivalence}) proves the other main result
of this section:
\begin{theorem}
  Solving \csproblem{Markovian Channel} or \csproblem{Markovian Map} is
  equivalent to solving P=NP: an efficient algorithm for
  \csproblem{Markovian Channel} or \csproblem{Markovian Map} would imply
  P=NP; conversely, P=NP would imply existence of efficient algorithms
  for \csproblem{Markovian Channel} and \csproblem{Markovian Map}.
\end{theorem}

\section{The Classical Problem}
\label{sec:classical}
The classical analogue of the Markovianity problem is called the
\emph{embedding problem}, but it is much older, dating back to at least
1937~\cite{Elfving}. For a given stochastic matrix $P$, the problem is to
determine whether or not $P$ can be embedded into a continuous-time
Markov chain, i.e.\ whether it is a member of a continuous-time,
one-parameter semigroup of stochastic matrices. Equivalently, does there
exist a generator $Q$ such that $P=e^Q$ and $e^{Qt}$ is stochastic for
all $t\geq 0$?

There is a long literature on the embedding problem, of which we do not
presume to give a comprehensive account here. (See \cite{embedding} for a more
extended history.) Simple necessary and sufficient conditions can easily be
derived for $2\times 2$ stochastic matrices (this result seems to originally
have been reported by Kingman~\cite{Kingman}, who attributes it to Kendall),
the $3\times 3$ case was eventually
solved~\cite{Cuthbert3x3,Johansen,Carette}, and certain properties are known
for the general case~\cite{Kingman_Williams,Fuglede,Cuthbert}. However, the
problem has remained open in general until now~\cite{Mukherjea,Davies}.

In order to discuss the complexity of the problem in a rigorous sense, it
is necessary to formulate the embedding problem as a weak-membership
problem, analogous to \csproblem{Markovian channel} or
\csproblem{Markovian map}, for the same reasons discussed in
\cref{sec:computational_Markovianity} in relation to the quantum
problem:
\begin{problem}[Embeddability]\label{prob:embedding}
  \begin{instance}
    $(P,\varepsilon)$: Stochastic matrix $P$; precision $\varepsilon \geq 0$.
  \end{instance}
  \begin{question}
    Assert either that:
    \begin{itemize}
    \item for some matrix $P'$ with $\matnorm{P'-P}\leq\varepsilon$, there
      exists a generator $Q'$ such that $P'=e^{Q'}$ and $e^{Q't}$ is
      stochastic for all $t\geq 0$;
    \item for some stochastic matrix $P'$ with
      $\matnorm{P'-P}\leq\varepsilon$, no such $Q'$ exists.
    \end{itemize}
  \end{question}
\end{problem}
Again, we could also formulate a variant analogous to
\csproblem{Markovian map}, which drops the requirement that the given $P$
be stochastic.

Now, stochastic maps are a special case of CPT maps in the following
sense. The diagonal entries of a density matrix form a probability
distribution, and every stochastic map can be extended to a CPT map whose
action on the subspace of diagonal density matrices is the same as the
action of the original stochastic map on the probability distribution
formed by those diagonal elements. For example, we can take the
composition of the CPT map that erases all off-diagonal elements of the
density matrix, with the original stochastic map acting on the diagonal
elements.

However, it does not follow that NP-hardness of the quantum problem
implies NP-hardness of the embedding problem, as that would require
precisely the opposite: encoding a CPT map into a stochastic map. But nor
would NP-hardness of the embedding problem imply NP-hardness of the
Markovianity problem, since the above argument showing that any
stochastic map can be extended to a CPT map does not ``preserve''
embeddability (more precisely, it does not map the set of stochastic maps
into the set of Markovian CPT maps, and the set of non-embeddable maps
into the set of non-Markovian CPT maps). The embedding problem for
stochastic matrices and the Markovianity problem for CPT maps are
inequivalent problems, and the complexity of each must be resolved
separately.

Fortuitously, it turns out that a proof of NP-hardness for the embedding
problem is already ``buried'' within the NP-hardness proof for the
Markovianity problem. We now give a sketch of the reduction from the
NP-complete \csproblem{1-in-3SAT} problem to the
\csproblem{Embeddability} problem of \cref{prob:embedding}, which closely
follows the analogous reduction to \csproblem{Markovian map}. For a full
account, see Ref.~\cite{embedding}.

Recall the conditions for $Q$ to be a generator of a continuous-time
Markov chain (a \keyword{$Q$-matrix}): (i) $Q_{i\neq j} \geq 0$, (ii)
$\sum_i Q_{i,j} = 0$. Comparing these with the conditions in
\cref{lem:L0+Ac_properties,eq:special_case_ccp1,eq:special_case_ccp2}
satisfied by $Q$ and $B^c$ from
\cref{eq:special_case_Q,eq:special_case_Bc}, we see that $Q_m = Q+2\pi
m_cB^c$ always satisfy the normalisation condition (ii) for any integers
$m_c$. But, from \cref{eq:special_case_ccp1} and the discussion
thereafter, $Q_m$ will satisfy condition (i) for some $m_c$ iff the
original \csproblem{1-in-3SAT} used to construct $Q$ and $B^c$ was
satisfiable. In other words, there exist integers $m_c$ such that $Q_m$
is a $Q$-matrix iff the \csproblem{1-in-3SAT} problem was satisfiable.
But $Q_m$ parametrise logarithms of the same matrix $P=e^{Q_m}$. In fact,
the only branches of the logarithm that are missing are branches that
could never generate a continuous-time Markov chain in any case. So,
either $P$ is not stochastic (which can easily be checked), in which case
the \csproblem{1-in-3SAT} problem cannot be satisfiable, or $P$ is
stochastic, in which case it is embeddable iff the \csproblem{1-in-3SAT}
problem was satisfiable.

To make this reduction rigorous,
\cref{lem:exp_continuity,cor:log_continuity} must be applied in very much
the same way as in the reduction from \csproblem{Lindblad generator} to
\csproblem{Markovian map} in \cref{thm:Lindblad-Markov_reduction}, to
show that a weak-membership formulation of the $Q$-matrix problem can be
reduced to the weak-membership formulation of the
\csproblem{Embeddability} problem (\cref{prob:embedding}). (See
Ref.~\cite{embedding} for a detailed treatment.) Similar arguments to
those given at the end of \cref{sec:NP-hardness} show that the
generalisation of the embedding problem to the problem of determining
whether a family of stochastic matrices are all generated by the same
continuous-time Markov process is also NP-hard, for any number of
matrices. Finally, it is clear how to adapt the algorithm of
\cref{sec:fixed_dimension} to the classical embedding problem, thereby
proving equivalence to P$=$NP.

\section{Conclusions}
\label{sec:conclusions}
We have shown that the Markovianity problem for CPT maps and the
analogous embedding problem for stochastic matrices are both NP-hard and,
indeed, have shown full equivalence between solutions to these problems
and a solution to the famous P$=$NP problem. Therefore, either P$=$NP, or
there exists no efficiently decidable criterion for deciding whether a
CPT map is generated by some underlying Markovian master equation, that
is, whether it is a member of a completely positive semi-group. Similarly
for deciding whether a stochastic matrix can be embedded in a
continuous-time homogeneous Markov process.

An interesting corollary of the NP-hardness proofs for the
\csproblem{Markovian channel} and \csproblem{Embeddability}
weak-membership problems is that:
\begin{corollary}\label{cor:non-zero_measure}
  Both the set of Markovian and the set of non-Markovian CPT maps have
  non-empty interior, hence non-zero measure, as do the sets of
  embeddable and non-embeddable stochastic matrices, in any finite
  dimension.
\end{corollary}
So a randomly chosen CPT map has a finite probability of being
non-Markovian, but also of being Markovian. The analogous property holds
for a randomly chosen stochastic map. Ref.~\cite{markovianity} estimates
these probabilities numerically for the simplest quantum case of qubits,
i.e.\ CPT maps on $\CC^2$. This fact alone may not be so surprising:
After all, generators being ccp can have neighbourhoods of generators that
are ccp, which under exponentiation are mapped to neighbourhoods of
channels, giving rise to a finite volume. The above corollary makes this
argument rigorous.

One consequence of these results to physics is that to decide whether a
given physical process at a shapshot in time---or for many snapshots for
that matter---is consistent with being forgetful cannot be decided
efficiently. This is because there is no \emph{a priori} way of knowing
whether the dynamics of an open system are Markovian or not, but finding
the dynamical equations (master equations) would answer this question,
and we now know this to be NP-hard for both the classical and quantum
cases, requiring infeasibly long computation time (unless P$=$NP, of
course). Whether this poses more practical difficulties is less
clear. The results of \cref{sec:fixed_dimension} show that it at least
does not pose a problem for the current generation of quantum
experiments, since other purely practical limitations on the dimension of
the systems being studied are more significant. More generally, one might
argue that the average-case complexity is more relevant in practice,
whereas NP-hardness only tells us about the worst-case complexity. What
is the average-case complexity of the Markovianity and embedding
problems? We close with this intriguing open problem, which we commend to
the reader.

\section{Acknowledgements}

The authors would like to thank Ignacio Cirac for numerous valuable
discussions relating to this work. TSC would like to thank Andreas Winter
for asking about classical analogues, and to an anonymous QIP conference
referee for pointing out a flaw in a previous treatment of the classical
case, which observation ultimately led to the NP-hardness proof for the
much older classical embedding problem. TSC also thanks Christina
Goldschmidt and James Martin for devoting time and patience to answering
his very basic questions about the relevant concepts in probability
theory. TSC was supported by a Leverhulme early career fellowship, and by
the European Commission QAP project, JE by the European Commission (QAP,
MINOS, COMPAS) and the EURYI, and MMW by QUANTOP and the Danish Research
Council.

\bibliographystyle{unsrt}
\bibliography{Markov_complexity}

\end{document}